\documentclass
[aps,prl,amsfonts,amssymb,twocolumn,amsmath,preprintnumbers,nofootinbib,floatfix,superscriptaddress]{revtex4-1}%
\usepackage[dvips]{graphics}
\usepackage{graphicx}
\usepackage{bm}
\usepackage{amsmath}
\usepackage{amsfonts}
\usepackage{amssymb}
\usepackage{xcolor}
\usepackage{subfigure}
\usepackage{tabularx}
\usepackage{multirow}
\usepackage{hyperref,hypcap}
\usepackage{braket}
\usepackage{commath}%
\providecommand{\U}[1]{\protect\rule{.1in}{.1in}}

\begin{document}

\title{Time-Reversal-Even Nonlinear Current Induced Spin Polarization}
\author{Cong Xiao}
\email{congxiao@hku.hk}
\thanks{These authors contributed equally to this work.}
\affiliation{Department of Physics, The University of Hong Kong, Hong Kong, People's Republic of China}
\affiliation{HKU-UCAS Joint Institute of Theoretical and Computational Physics at Hong Kong, People's Republic of China}
\author{Weikang Wu}
\thanks{These authors contributed equally to this work.}
\affiliation{Key Laboratory for Liquid-Solid Structural Evolution and Processing of Materials, Ministry of Education, Shandong University, Jinan 250061, People's Republic of China}
\author{Hui Wang}
\thanks{These authors contributed equally to this work.}
\affiliation{Research Laboratory for Quantum Materials, Singapore University of Technology and Design, Singapore 487372, Singapore}
\author{Yue-Xin Huang}
\thanks{These authors contributed equally to this work.}
\affiliation{Research Laboratory for Quantum Materials, Singapore University of Technology and Design, Singapore 487372, Singapore}
\author{Xiaolong Feng}
\affiliation{Research Laboratory for Quantum Materials, Singapore University of Technology and Design, Singapore 487372, Singapore}
\author{Huiying Liu}
\email{liuhuiying@pku.edu.cn}
\affiliation{Research Laboratory for Quantum Materials, Singapore University of Technology and Design, Singapore 487372, Singapore}
\author{Guang-Yu Guo}
\email{gyguo@phys.ntu.edu.tw}
\affiliation{Department of Physics, National Taiwan University, Taipei 10617, Taiwan, Republic of China}
\affiliation{Physics Division, National Center for Theoretical Sciences, Taipei 10617, Taiwan, Republic of China}
\author{Qian Niu}
\affiliation{School of Physics, University of Science and Technology of China, Hefei, Anhui 230026, People's Republic of China}
\author{Shengyuan A. Yang}
\affiliation{Research Laboratory for Quantum Materials, Singapore University of Technology and Design, Singapore 487372, Singapore}

\begin{abstract}
We propose a time-reversal-even spin generation in second order of electric
fields, which dominates the current induced spin polarization in a wide class
of centrosymmetric nonmagnetic materials, and leads to a novel nonlinear spin-orbit
torque in magnets. We reveal a quantum origin of this effect from the
momentum space dipole of the anomalous spin polarizability.
First-principles calculations predict sizable spin generations in several
nonmagnetic hcp metals, in monolayer TiTe$_{2}$, and in ferromagnetic monolayer
MnSe$_{2}$, which can be detected in experiment. Our work opens up the broad
vista of nonlinear spintronics in both nonmagnetic and magnetic systems.

\end{abstract}
\maketitle

Nonlinear responses of solids are attracting great interest in recent research \cite{Ma2021,Lu2021}. They dominate in crystals
where the linear response is symmetry forbidden, probe novel band geometric quantities, and offer new tools to characterize and control material properties. For example, recent studies on various nonlinear anomalous Hall effects have connected them to
intriguing geometric quantities such as Berry curvature dipole \cite{Fu2015,Ma2019,Kang2019} and Berry connection polarizability
\cite{Gao2014,Lai2021,Liu2022}, and revealed their utility to extract N\'{e}el vector \cite{Shao2020,Wang2021,Liu2021} and to detect nontrivial band topologies \cite{Facio2018,Amit2022}.

In the field of spintronics, current induced spin polarization (CISP) is the central effect that enables electric control of spin degree of freedom \cite{Review2009,Manchon2019}. In this context, one usually distinguishes contributions according to their parities under time reversal ($\mathcal{T}$), i.e., under the reversal of all magnetic moments in the system \cite{Freimuth2014,Zelezny2017,Manchon2019}. Clearly, the $\mathcal{T}$-odd part is present only in magnets, whereas the $\mathcal{T}$-even CISP exists in both magnetic and nonmagnetic systems \cite{Pikus1978,Aronov1989,Edelstein,Culcer2007}. It was shown that they give rise to two basic types of spin-orbit torques, allowing electrical manipulation of magnetic order parameters \cite{Geller2009,Garate2009,Franz2010,Miron2010,Vyborny2011,Miron2011,Liu2012,Garello2013,Kurebayashi2014,Jungwirth2016,Manchon2019}. Previous studies focused on the linear CISP, which is limited to systems with inversion symmetry ($\mathcal{P}$) breaking. This left out the large family of $\mathcal{P}$ symmetric materials, in which the leading CISP is necessarily of nonlinear character. Recently, the $\mathcal{T}$-odd second-order nonlinear CISP was proposed in Ref. \cite{Xiao2022NLISOT}. However, the corresponding $\mathcal{T}$-even part has not been investigated yet. As mentioned, the $\mathcal{T}$-even nonlinear CISP occurs in even wider range of material systems, including also the nonmagnetic materials, many of which are technologically important (such as the elemental metals).

In this work, we study a special type of such a $\mathcal{T}$-even nonlinear CISP. It has a quantum origin arising
from the \textit{anomalous spin}, which is spotlighted here as a basic
property of spin-orbit-coupled electrons under an electric field and can be expressed in terms of an intrinsic band geometric quantity which is called the anomalous spin polarizability (ASP).
We show that the $\mathcal{T}$-even nonlinear spin response is determined by the momentum space dipole of ASP over the occupied states.
We clarify the symmetry character of this effect and find that in several magnetic crystal classes, the $\mathcal{T}$-even and $\mathcal{T}$-odd contributions give orthogonal spin polarizations, hence their effects can be readily separated in experiment.
Combining our theory with first-principles calculations, we report sizable nonlinear spin generations in a number of nonmagnetic elemental metals, in monolayer TiTe$_{2}$, and in ferromagnetic monolayer MnSe$_{2}$. Our finding establishes the $\mathcal{T}$-even nonlinear CISP as a fundamental spintronic effect, which renders a new nonlinear spin-orbit torque in centrosymmetric magnets.


\emph{{\color{blue} Symmetry characters.}} Let us first understand the emergence
of $\mathcal{T}$-even nonlinear CISP from the symmetry perspective. The
quadratic spin polarization response $\delta \bm s$ to an applied electric field can be expressed as
\begin{equation}
\delta s_{a}=\alpha_{abc}E_{b}E_{c}, \label{1}%
\end{equation}
where $\alpha$ is the nonlinear response tensor, the roman indices label the Cartesian components, and the Einstein summation convention is adopted. Obviously, in a $\mathcal{P}$-symmetric system, the linear response $\delta s\propto E$ is forbidden, and (\ref{1}) becomes the leading effect.

$\alpha_{abc}$ can always be separated into a
$\mathcal{T}$-even part and a $\mathcal{T}$-odd part: $\alpha=\alpha^{\text{even}}+\alpha^{\text{odd}}$. Note that in nonmagnetic materials, only $\alpha^{\text{even}}$ exists. The two parts have different symmetry properties. For $\alpha^{\text{odd}}$, these can be found in \cite{Xiao2022NLISOT}. Here, we focus on $\alpha^{\text{even}}$, which obeys the following symmetry transformation rule:
\begin{equation}
\alpha_{a^{\prime}b^{\prime}c^{\prime}}^{\text{even}}=\text{det}%
(\mathcal{R})\mathcal{R}_{a^{\prime}a}\mathcal{R}_{b^{\prime}b}\mathcal{R}%
_{c^{\prime}c}\alpha_{abc}^{\text{even}}, \label{eq:constraints}%
\end{equation}
{with }$\mathcal{R}$ being a point group operation.  The obtained constraints are summarized in Table~\ref{operations}.


From the analysis, we find that the
$\mathcal{T}$-even nonlinear CISP is supported by 10 of the 11 centrosymmetric
point groups, implying broad material platforms in which the effect could be dominating. The detailed forms of
$\alpha_{abc}^{\text{even}}$ tensor constrained by symmetry are presented in
the Supplemental Material \cite{supp}. Importantly, we find that in $\bar{6}m'2'$, $4/mm'm'$, $6/mm'm'$ and $\bar{3}m'$ magnetic groups, which do not support the linear CISP, the nonlinear CISPs due to $\mathcal{T}$-even and $\mathcal{T}$-odd parts must be along orthogonal directions for any direction of the driving electric field \cite{supp}, thus allowing an easy separation of the two parts.

\renewcommand{\arraystretch}{1.42} \begin{table}[ptb]
\caption{Constraints on $\alpha_{a\left(  bc\right)  }^{\text{even}}%
=(\alpha_{abc}^{\text{even}}+\alpha_{acb}^{\text{even}})/2$ from magnetic
point group symmetries. \textquotedblleft$\checkmark$\textquotedblright%
\ (\textquotedblleft$\times$\textquotedblright) means that the element is
symmetry allowed (forbidden). Symmetry operations $\mathcal{RT}$ and
$\mathcal{R}$ impose the same constraints. For simplicity, we assume the $E$
field is applied within the $xy$ plane. }%
\label{operations}%
\begin{centering}
\begin{tabular}{p{0.12\linewidth}p{0.09\linewidth}<{\centering}p{0.09\linewidth}<{\centering}
p{0.09\linewidth}<{\centering}p{0.09\linewidth}<{\centering}p{0.09\linewidth}<{\centering}
p{0.09\linewidth}<{\centering}p{0.09\linewidth}<{\centering}p{0.09\linewidth}<{\centering}}
\hline \hline
& \multirow{2}{*}{$\mathcal{P}$} & \multirow{2}{*}{$C_{2}^{z}$} & $C_{3}^{z}$, & $C_{4,6}^{z}$, & $C_{2,4,6}^{x}$, & $C_{3}^{x}$, & \multirow{2}{*}{$\sigma_{z}$} & \multirow{2}{*}{$\sigma_{x}$}\tabularnewline[-1.2ex]
& & &$S_{6}^{z}$  & $S_{4}^{z}$ & $S_{4}^{x}$ & $S_{6}^{x}$ & & \tabularnewline\hline
\ $\alpha^{\text{even}}_{xxx}$ & $\checkmark$ & $\times$ & $-\alpha^{\text{even}}_{xyy}$ & $\times$ & $\checkmark$ & $\checkmark$ & $\times$ & $\checkmark$\tabularnewline
\ $\alpha^{\text{even}}_{x(xy)}$ & $\checkmark$ & $\times$ & $\alpha^{\text{even}}_{yxx}$ & $\times$ & $\times$ & $\times$ & $\times$ & $\times$ \tabularnewline
\ $\alpha^{\text{even}}_{xyy}$ & $\checkmark$ & $\times$ & $\checkmark$ & $\times$ & $\checkmark$ & $\checkmark$ & $\times$ & $\checkmark$ \tabularnewline
\ $\alpha^{\text{even}}_{yxx}$ & $\checkmark$ & $\times$ & $\checkmark$ & $\times$ & $\times$ & $\times$ & $\times$ & $\times$
\tabularnewline
\ $\alpha^{\text{even}}_{y(xy)}$ & $\checkmark$ & $\times$ & $\alpha^{\text{even}}_{xyy}$ & $\times$ & $\checkmark$ & $\checkmark$ & $\times$ & $\checkmark$ \tabularnewline
\ $\alpha^{\text{even}}_{yyy}$ & $\checkmark$ & $\times$ & $-\alpha^{\text{even}}_{yxx}$ & $\times$ & $\times$ & $\checkmark$ & $\times$ & $\times$ \tabularnewline
\ $\alpha^{\text{even}}_{zxx}$ & $\checkmark$ & $\checkmark$ & $\checkmark$ & $\checkmark$ & $\times$ & $\times$ & $\checkmark$ & $\times$\tabularnewline
\ $\alpha^{\text{even}}_{z(xy)}$ & $\checkmark$ & $\checkmark$ & $\times$ & $\times$ & $\checkmark$ & $\checkmark$ & $\checkmark$ & $\checkmark$ \tabularnewline
\ $\alpha^{\text{even}}_{zyy}$ & $\checkmark$ & $\checkmark$ & $\alpha^{\text{even}}_{zxx}$ & $\alpha^{\text{even}}_{zxx}$ & $\times$ & $\checkmark$ & $\checkmark$ & $\times$ \tabularnewline
\hline \hline
\end{tabular}
\par\end{centering}
\end{table}

\emph{{\color{blue} ASP dipole mechanism.}} The spin density is given by the
integral of the spin polarization $\bm s^{n}(\bm k)$ carried by each electron wave packet
weighted by the distribution function $f_{n}(\bm k)$ (we set $e=\hbar=1$):
\begin{equation}
\bm s=\int[d\bm k]f_{n}(\bm k)\bm s^{n}(\bm k), \label{S}%
\end{equation}
where $n$ and $\bm k$ are the band index and the wave vector,
respectively, and $[d\bm k]$ is shorthand for $\sum_{n}d\bm k/(2\pi)^{d}$ with
$d$ being the dimension of the system. The $\mathcal{T}$-even response
requires a distribution function that breaks the occupation symmetry at $\bm k$ and $-\bm k$, otherwise it
would vanish in nonmagnetic systems due to the Kramers degeneracy.
This is provided by the nonequilibrium distribution computed to the first order of the driving $E$ field. Using the Boltzmann equation with the constant relaxation time approximation, we have
$f_{n}=f_{0}-\tau E_{c}\partial_{c}f_{0}$, where $f_{0}$ is the equilibrium Fermi distribution, $\tau$ is the relaxation time, and $\partial_{c}\equiv\partial_{k_{c}}$. Meanwhile, with spin-orbit coupling,
 $\bm s^{n}(\bm k)$ of a wave packet also acquires a correction by the $E$ field \cite{Dong2020}:
\begin{equation}
s_{a}^{n}(\bm k)=\langle u_{n}(\bm k)|\hat{s}_{a}|u_{n}(\bm k)\rangle
+\Upsilon_{ab}^{n}(\bm k)E_{b}, \label{Abelian}%
\end{equation}
where the first term is the expectation value of the spin operator for the
eigenstate $|u_{n}(\bm k)\rangle$, and the second term is the \emph{anomalous
spin} correction linear in the electric field. The coefficient of the correction,
\begin{equation}
\Upsilon_{ab}^{n}(\bm k)=2\operatorname{Im}\sum_{n^{\prime}\neq n}\frac{s_{a}^{nn^{\prime}}(\bm k)v_{b}%
^{n^{\prime}n}(\bm k)}{\big[\varepsilon_{n}(\bm k)-\varepsilon_{n^{\prime}%
}(\bm k)\big]^{2}}, \label{geometric}%
\end{equation}
is the ASP, an intrinsic band geometric quantity representing the polarizability of anomalous spin to the applied $E$ field. In (\ref{geometric}),
$\varepsilon_{n}(\bm k)$ is the band energy, and the numerator involves the interband matrix elements of
spin and velocity operators.

Substituting the expressions of $f_n$ and (\ref{Abelian}) into (\ref{S}) and collecting the $\mathcal{T}$-even terms of $E^{2}$ order,
we obtain the nonlinear CISP response tensor
\begin{equation}
  \alpha_{abc}^{\text{even}}=\tau \mathcal{D}_{abc},
\end{equation}
with%
\begin{equation}
\mathcal{D}_{abc}=\int[d\bm k]f_{0}\partial_{c}\Upsilon_{ab}%
^{n}\label{both}%
\end{equation}
being the momentum space dipole moment of ASP over all occupied states in equilibrium.
One checks that this is indeed a
$\mathcal{T}$-even pseudotensor complying with the symmetry analysis. It is a Fermi surface property
as can be seen via an integration by
parts in (\ref{both}).


Our result shows that the $\mathcal{T}$-even nonlinear CISP is proportional to the ASP dipole. This is
analogous to the $\mathcal{T}$-even nonlinear anomalous Hall effect in Ref. \cite{Fu2015}, which is proportional to the Berry curvature dipole. In fact, the expression of ASP [Eq. (\ref{geometric})] is also similar to the Berry curvature tensor $\Omega_{ab}$ \cite{Xiao2010}, with one of the velocity matrix element replaced by the spin matrix element. The analogy can be further exemplified by comparing Eq.~(\ref{Abelian}) with the well-known semiclassical equation of motion: $\dot{r}^n_a=v^{nn}_a+\Omega_{ab}E_b$ \cite{Chang1995,Sundaram1999,Xiao2010}. One directly observes that the anomalous spin parallels the anomalous velocity $\Omega_{ab}E_b$. Moreover, while integrating the anomalous velocity over occupied states gives the intrinsic linear anomalous Hall effect, the integration of anomalous spin also produces the intrinsic linear CISP \cite{Garate2009,Franz2010,Kurebayashi2014}. This highlights the significance of anomalous spin (and ASP) as an essential ingredient in the description of spin-orbit-coupled Bloch electrons.

\begin{figure}[t]
\centering
\includegraphics[width=0.45\textwidth]{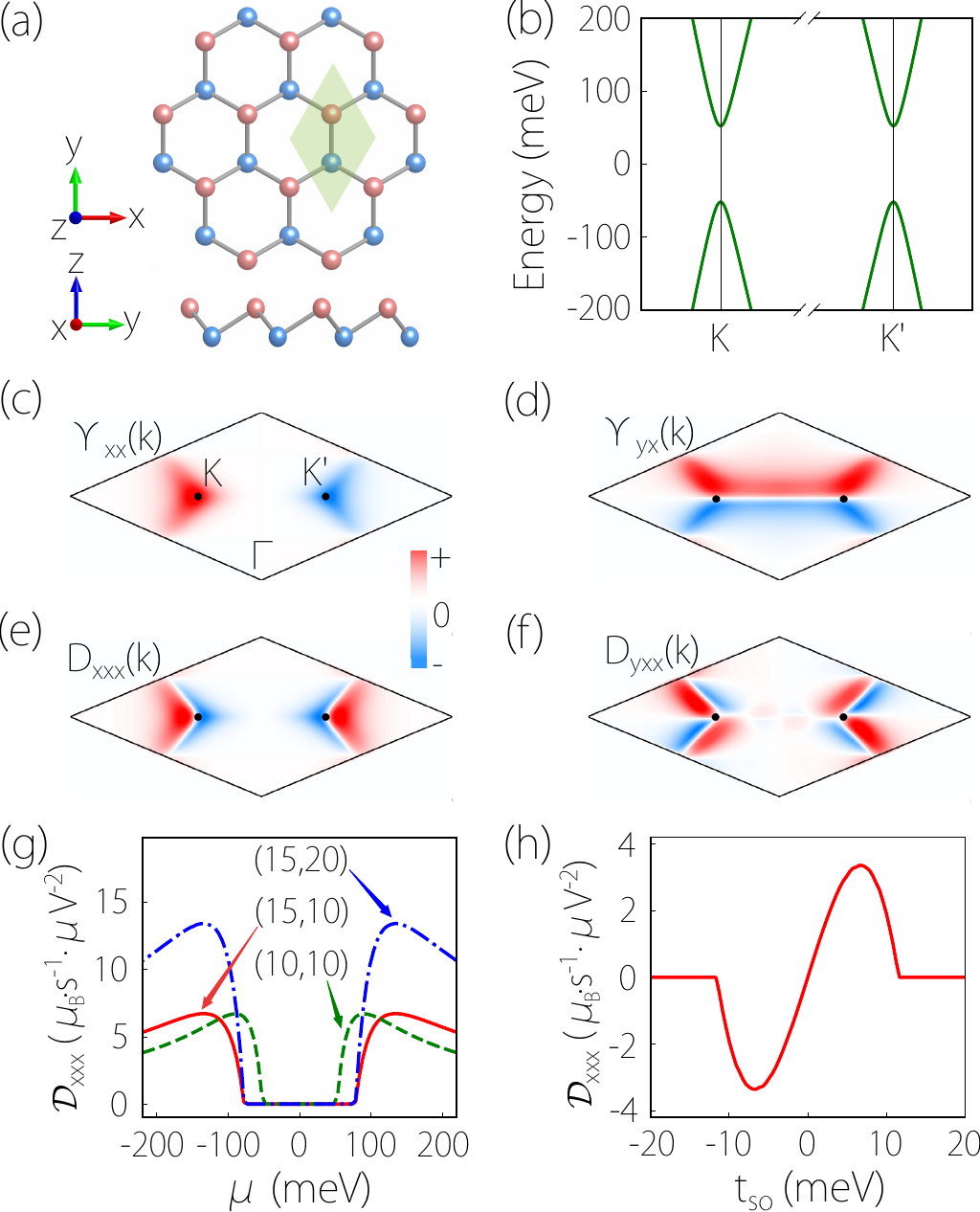} \caption{(a) Top and
side views of the buckled honeycomb lattice for the modified Kane-Mele model.
(b) Low-energy  bands for this model, with two valleys at $K$ and $K'$.
Here, we take $t=0.85$ eV, $t_{so}=10$ meV, and $t_{R}=10$ meV.
(c-f) show the $k$-space distribution of the ASP (c) $\Upsilon_{xx}$, (d) $\Upsilon_{yx}$ and the $k$-resolved ASP dipole (e) $D_{xxx}(\boldsymbol{k})$ and (f) $D_{yxx}(\boldsymbol{k})$ for the valence band.
(g) Calculated ASP dipole $\mathcal{D}_{xxx}$ versus the chemical potential $\mu$.
The legends denote $(t_{so},t_{R})$ in units of meV. (h) $\mathcal{D}_{xxx}$ versus $t_{so}$, for $\mu=-60$ meV and $t_{R}=10$ meV.}%
\label{fig-stanene}%
\end{figure}

\emph{{\color{blue} A model study.}} To illustrate the features of ASP dipole
and the resulting CISP, we first apply our theory to a modified Kane-Mele
model defined on a buckled two-dimensional (2D) honeycomb
lattice [Fig.~\ref{fig-stanene}(a)] \cite{Kane2005,Yao2011PRL,Yao2011PRB}, which reads
\begin{align}
H    =&-t\sum_{\left\langle ij\right\rangle \sigma}c_{i\sigma}^{\dag
}c_{j\sigma}+it_{so}\sum_{\langle\langle ij\rangle\rangle\sigma\sigma^{\prime
}}\nu_{ij}c_{i\sigma}^{\dag}s_{\sigma\sigma^{\prime}}^{z}c_{j\sigma^{\prime}%
}\nonumber\\
&  -it_{R}\sum_{\left\langle \left\langle ij\right\rangle \right\rangle
\sigma\sigma^{\prime}}\mu_{ij}c_{i\sigma}^{\dag}(\boldsymbol{s}\times
\boldsymbol{d}_{ij})_{\sigma\sigma^{\prime}}^{z}c_{j\sigma^{\prime}}.
\label{model}%
\end{align}
Here $c_{i\sigma}$ ($c_{i\sigma}^{\dagger}$) is the annihilation (creation)
operator for an electron with spin $\sigma$ at site $i$. The first term is the
nearest neighbor hopping. The second term is the intrinsic spin-orbit coupling
in second neighbor hopping, where $\nu_{ij}=+ (-)$ if the electron makes a left (right) turn during hopping from
$j$ to $i$.
The third term is the intrinsic Rashba spin-orbit coupling due to lattice buckling, where $\boldsymbol{d}%
_{ij}$ is the unit vector pointing from site $j$ to $i$. This term is needed to lower the symmetry from $D_{6h}$ to $D_{3d}$, such that a nonlinear spin polarization can be induced by an \emph{in-plane} $E$ field.

According to
Table~\ref{operations}, the symmetries $C_{3}^{z}$, $C_{2}^{x}$, and
$\sigma_{x}$ of $D_{3d}$ group enforce the following relations on $\alpha^\text{even}$: $\alpha_{xyy}^{\text{even}}=\alpha_{y(xy)}%
^{\text{even}}=-\alpha_{xxx}^{\text{even}}$. It follows that the system only
allows an in-plane spin polarization in the form of
\begin{equation}
(\delta s_{x},\delta s_{y})=\alpha_{xxx}^{\text{even}}(\cos2\phi,-\sin
2\phi)E^{2}. \label{D3d}%
\end{equation}
Interestingly, the result is determined by a single independent element $\alpha_{xxx}^{\text{even}}$, and
exhibits an angular dependence with $\pi$ periodicity. Here, $\phi$ is the
polar angle of the in-plane $E$ field measured from the $C_{2}^{x}$ axis.

In Figs.~\ref{fig-stanene}(c) -- (f), we plot in the Brillouin zone the distribution of ASP as well as the $k$-resolved ASP dipole, i.e.,
$D_{abc}(\boldsymbol{k})=\sum_{n}f_{0}\partial_{c}\Upsilon_{ab}^{n}$, the integrand of (\ref{both}). One observes that these quantities are
concentrated around the small-gap region in the band structure, reflecting the interband coherence
nature of band geometric quantities. Here, $D_{yxx}(\boldsymbol{k})$ is odd in
$k_{y}$, whereas $D_{xxx}(\boldsymbol{k})$ is even, resulting in a non-vanishing ASP dipole
$\mathcal{D}_{xxx}$. In Fig.~\ref{fig-stanene}(g), we plot $\mathcal{D}_{xxx}$ versus the chemical potential $\mu$, which shows that the ASP dipole is enhanced around the band edges.
In Fig.~\ref{fig-stanene}(h), we further see that $\mathcal{D}_{xxx}$ flips its sign with the
spin-orbit coupling $t_{so}$.


\begin{table}[ptb]
\caption{Calculated ASP dipole and CISP of some hcp transition metals at room
temperature (RT). The RT transport relaxation time $\tau$ is obtained by using
experimental resistivity data \cite{Kittel1996} and calculated Drude weight.
The driving electric field is taken as $E=10^{5}$ V/m \cite{Geller2009,Jungwirth2018,Song2019}.}
\label{hcp}%
\begin{ruledtabular}
\begin{tabular}{c c c c}
System    & $\mathcal{D}_{y(zx)}$                                   & $\delta s_{y}$   & $\tau$ \\
& $10^{18}$ ($\mu_B$/cm$^3$)[s(V/m)$^2$]$^{-1}$ & $10^{-7}$ ($\mu_B$/nm$^3$) & (10 fs) \\ \hline
Ti    & -0.50  &-0.43 & 0.85 \\
Zr    &  1.69  & 1.34 & 0.79 \\
Hf    & -0.73  &-0.80 & 1.10 \\
Re    & -0.93  &-0.60 & 0.64 \\
Ru    & -1.18  &-1.00 & 0.85 \\
\end{tabular}
\end{ruledtabular}
\end{table}

\emph{{\color{blue} Application to nonmagnetic metals.}} Next, we ask if the
$\mathcal{T}$-even nonlinear CISP is appreciable in real materials.
Combining our theory with first-principles calculations, we first evaluate the effect in several hcp
nonmagnetic transition metals, as listed in Table \ref{hcp}, which have served as a main platform for studying
the spin Hall effect \cite{Freimuth2010}. These materials forbid the linear CISP because of
their inversion symmetry.

Take the $x/z$ to be along the crystal $a/c$
axis, the $D_{6h}$ point group dictates only one
nonzero independent element of ASP dipole $\mathcal{D}_{y(xz)}$, and the induced
spin polarization must take the form of%
\begin{equation}
\delta\boldsymbol{s}=\alpha_{y\left(  xz\right)  }^{\text{even}}\sin
2\theta\left(  -\sin\phi,\cos\phi,0\right)  E^{2},%
\end{equation}
lying in the $xy$ plane, where $\theta$ and $\phi$ are spherical
angles for the $E$ field.
Notably, $\delta\boldsymbol{s}$ is always
normal to the field, and has a $2\pi$ periodicity in $\phi$ and a $\pi$
periodicity in $\theta$. For instance, if the $E$ field is applied within the
$zx$ plane, the induced spin is along the $y$ axis and reads $\delta
s_{y}=\alpha_{y\left(  xz\right)  }^{\text{even}}\sin2\theta E^{2}$, which
reaches its maximum magnitude when $\theta=\pi/4$ and $3\pi/4$.

The results of $\mathcal{D}_{y(xz)}$ and $\delta s_{y}$ from our first-principles calculations (calculation
details in \cite{supp}) are shown in Table~\ref{hcp}. The
induced spin density can reach $10^{-7} \mu_{B}/\mathrm{nm^{3}}$ at $E=10^5$ V/m, which is considerable
compared to the linear CISP ($\sim 10^{-9}$ to $10^{-8} \mu_{B}/\mathrm{nm^{3}}$) that has been measured in noncentrosymmetric
nonmagnetic systems in previous experiments \cite{Kato2004,Stern2006}.
In the Supplemental Material \cite{supp}, we also show the result for a 2D $\mathcal{P}$-symmetric nonmagnetic metal,
the experimentally synthesized monolayer TiTe$_{2}$ \cite{Chiang2017}, which shows nonlinear CISP of a similar magnitude.

\begin{figure}[ptb]
\centering
\includegraphics[width=0.49\textwidth]{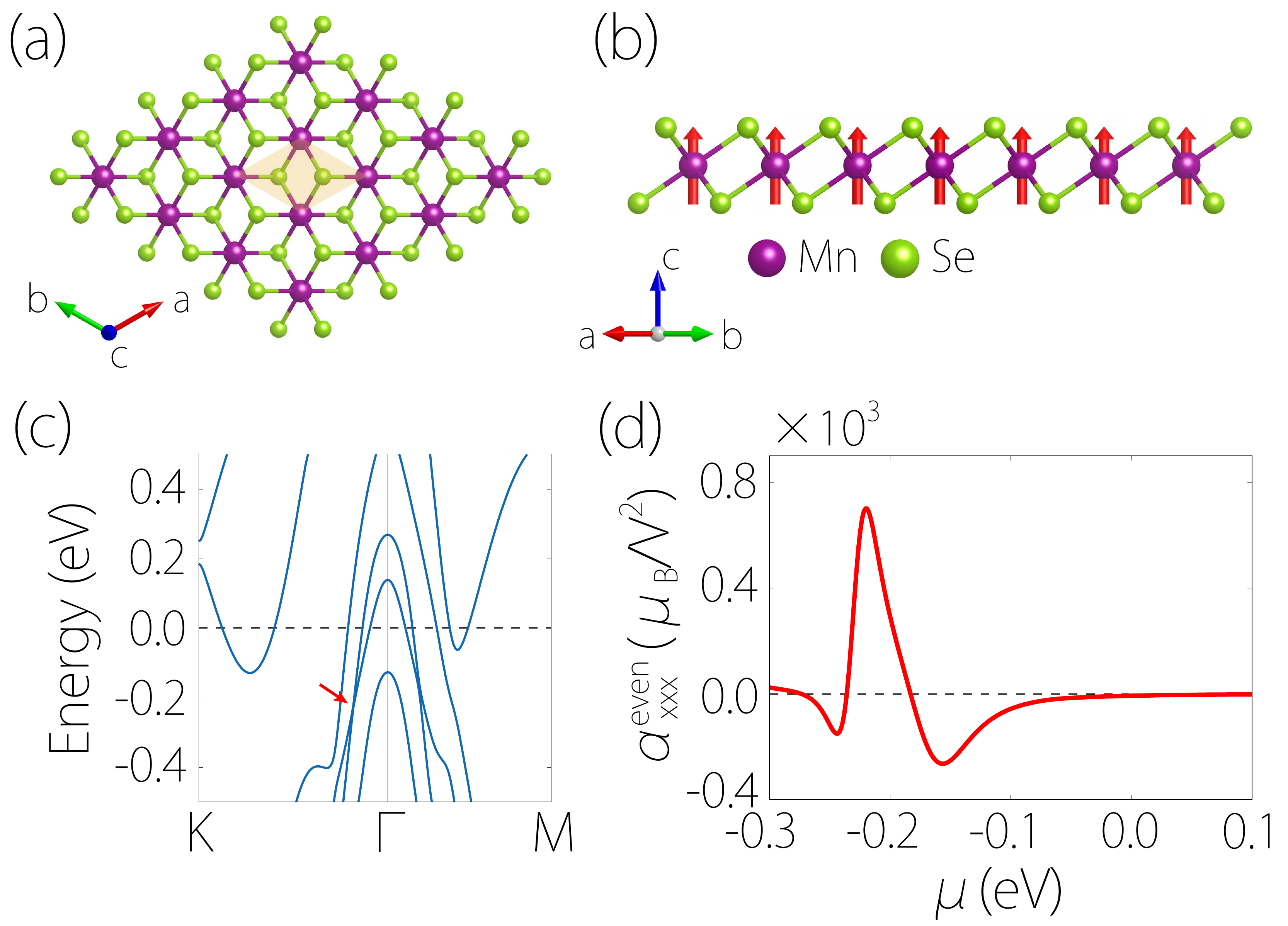} \caption{(a) Top and
(b) side views of the structure of monolayer MnSe$_{2}$. (c) Calculated band
structure for monolayer MnSe$_{2}$ in the ferromagnetic phase.
(d) Calculated $\alpha_{xxx}^{\text{even}}$ versus the chemical potential.}%
\label{sfig:MnSe2}%
\end{figure}

{\color{blue} \emph{Application to ferromagnetic 2D MnSe$_{2}$.}}
Our second example is the monolayer 1T-MnSe$_{2}$, which has been synthesized in recent experiment and demonstrated to be a
room-temperature 2D ferromagnetic metal \cite{Kawakami2018,Magnus2020}. Its
lattice structure is shown in Figs.~\ref{sfig:MnSe2}(a) and (b), where each Mn
atom is located at an inversion center, and the lattice point group is
$D_{3d}$.
Previous experiment \cite{Kawakami2018} showed that the magnetization is out-of-plane (along $z$), so the magnetic point
group is $\bar{3}m^{\prime}$. Again, in this system, the linear CISP
is forbidden by $\mathcal{P}$, thus the nonlinear effect dominates. The CISP
constrained by symmetry takes the same form as Eq.~(\ref{D3d}). Thus, for an
in-plane $E$ field, the CISP is also in-plane and hence normal to the
equilibrium magnetization.


The calculated band structure is plotted in Fig.~\ref{sfig:MnSe2}(c). Figure~\ref{sfig:MnSe2}(d) shows the
variation of $\alpha_{xxx}^{\text{even}}$ with respect to the Fermi energy
$\mu$. As there has been no reported values for the relaxation time in
monolayer MnSe$_{2}$, we take $\tau$ as 0.05~ps, a typical value for 2D
metals at low temperatures \cite{Ma2019}. Then, $\alpha_{xxx}%
^{\text{even}}$ is found to be $\sim -7.1$~$\mu_{B}$/V$^{2}$ without doping, and is
greatly enhanced upon hole doping, reaching $\sim
7\times10^{2}$~$\mu_{B}$/V$^{2}$ at $\mu=-0.22$~eV as a result of the small
local gap (about 15 meV), as marked by the red arrow in
Fig.~\ref{sfig:MnSe2}(c). In practice, such a doping level can be achieved in 2D materials by
electric gating \cite{Xie2010,Ma2019}.

Under a moderate driving field of $10^{5}$~V/m \cite{Geller2009,Jungwirth2018,Song2019},
the $\mathcal{T}$-even nonlinear CISP can reach
$\sim0.7\times10^{-5}$~$\mu_{B}$/nm$^{2}$ (or $2.5\times10^{-5}$~$\mu_{B}%
$/nm$^{3}$ considering the monolayer thickness). Previous experiments showed that
the linear CISP with much smaller magnitude, e.g., $\sim10^{-9}$ to $10^{-6}$ $\mu
_{\text{B}}$/nm$^{3}$, can be measured in ferromagnets by magneto-optical or anisotropic magnetoresistance
effects and can drive magnetization dynamics \cite{Geller2009,Vyborny2011,Kurebayashi2014}. Thus, the predicted effect here is indeed
significant. It should be readily detectable and can produce sizable spin-orbit torques.


In magnets, the $\mathcal{T}$-odd nonlinear CISP may also be present \cite{Xiao2022NLISOT},
which can be at the zeroth or the second order of $\tau$.
In experiment, such contributions can be distinguished from the $\mathcal{T}$-even effect here by their different $\tau$ scaling (e.g., by plotting against the longitudinal conductivity with varying temperature) \cite{Kang2019,Lai2021}.
Another way to separate them is to utilize their different symmetry properties. As mentioned, for the $\bar{3}m^{\prime}$ group, to which the ferromagnetic 2D MnSe$_{2}$ belongs, the $\mathcal{T}$-even and $\mathcal{T}$-odd CISPs are always orthogonal (see the Supplemental Material \cite{supp}). For example, with $E$ field along $x$, the $\mathcal{T}$-even CISP is along $x$, whereas the $\mathcal{T}$-odd CISP is in the $yz$ plane. This permits an easy separation of the two effects \cite{Manchon2019}.


\emph{{\color{blue} Discussion.}} We have proposed the $\mathcal{T}$-even nonlinear CISP effect and revealed its geometric origin in the ASP dipole. It offers a new mechanism for electric control of
spin in $\mathcal{P}$-symmetric nonmagnetic materials and for driving spin-orbit torques in ferromagnets.
We demonstrate the first-principles evaluation
of the effect for concrete nonmagnetic and ferromagnetic materials, with
sizable results well within the capacity of experiments. The study can be
naturally extended to other materials.
In addition, the effect is also expected to
play a role in $\mathcal{P}$-broken systems if the inversion symmetry is not
strongly broken.


As mentioned, the ASP dipole mechanism for the nonlinear CISP parallels the Berry curvature dipole mechanism
in the nonlinear anomalous Hall effect \cite{Fu2015}. The
relaxation time approximation is adopted here, so that the detailed disorders' forms which are usually
unknown do not pose a difficulty. Contributions beyond this approximation
can be approached via semiclassical or quantum kinetic theories by assuming some
specific form of disorder, in a way parallel to the study of nonlinear
Hall effect \cite{Du2019,Xiao2019NLHE}.



We have not addressed the effect in antiferromagnets, because the relation
between CISP and N\'{e}el spin-orbit torque is different from that in
ferromagnets. Consider collinear antiferromagnets, it is the staggered CISP on
different magnetic sublattices that contributes to the N\'{e}el
torque \cite{Zelezny2014,Zelezny2017}. As such, the properties of the required nonlinear CISP must be analyzed from the magnetic \textit{space} groups rather than point groups. This deserves a
separate thorough study and will be important for the electrical control of
N\'{e}el vector in locally centrosymmetric (i.e., each magnetic sublattice preserves $\mathcal{P}$) antiferromagnets.


\begin{acknowledgments}
\end{acknowledgments}

\bibliographystyle{apsrev4-1}
\bibliography{NLSP_ref}

\begin{thebibliography}{53}%
\makeatletter
\providecommand \@ifxundefined [1]{%
 \@ifx{#1\undefined}
}%
\providecommand \@ifnum [1]{%
 \ifnum #1\expandafter \@firstoftwo
 \else \expandafter \@secondoftwo
 \fi
}%
\providecommand \@ifx [1]{%
 \ifx #1\expandafter \@firstoftwo
 \else \expandafter \@secondoftwo
 \fi
}%
\providecommand \natexlab [1]{#1}%
\providecommand \enquote  [1]{``#1''}%
\providecommand \bibnamefont  [1]{#1}%
\providecommand \bibfnamefont [1]{#1}%
\providecommand \citenamefont [1]{#1}%
\providecommand \href@noop [0]{\@secondoftwo}%
\providecommand \href [0]{\begingroup \@sanitize@url \@href}%
\providecommand \@href[1]{\@@startlink{#1}\@@href}%
\providecommand \@@href[1]{\endgroup#1\@@endlink}%
\providecommand \@sanitize@url [0]{\catcode `\\12\catcode `\$12\catcode
  `\&12\catcode `\#12\catcode `\^12\catcode `\_12\catcode `\%12\relax}%
\providecommand \@@startlink[1]{}%
\providecommand \@@endlink[0]{}%
\providecommand \url  [0]{\begingroup\@sanitize@url \@url }%
\providecommand \@url [1]{\endgroup\@href {#1}{\urlprefix }}%
\providecommand \urlprefix  [0]{URL }%
\providecommand \Eprint [0]{\href }%
\providecommand \doibase [0]{http://dx.doi.org/}%
\providecommand \selectlanguage [0]{\@gobble}%
\providecommand \bibinfo  [0]{\@secondoftwo}%
\providecommand \bibfield  [0]{\@secondoftwo}%
\providecommand \translation [1]{[#1]}%
\providecommand \BibitemOpen [0]{}%
\providecommand \bibitemStop [0]{}%
\providecommand \bibitemNoStop [0]{.\EOS\space}%
\providecommand \EOS [0]{\spacefactor3000\relax}%
\providecommand \BibitemShut  [1]{\csname bibitem#1\endcsname}%
\let\auto@bib@innerbib\@empty
\bibitem [{\citenamefont {Ma}\ \emph {et~al.}(2021)\citenamefont {Ma},
  \citenamefont {Grushin},\ and\ \citenamefont {Burch}}]{Ma2021}%
  \BibitemOpen
  \bibfield  {author} {\bibinfo {author} {\bibfnamefont {Q.}~\bibnamefont
  {Ma}}, \bibinfo {author} {\bibfnamefont {A.~G.}\ \bibnamefont {Grushin}}, \
  and\ \bibinfo {author} {\bibfnamefont {K.~S.}\ \bibnamefont {Burch}},\ }\href
  {\doibase 10.1038/s41563-021-00992-7} {\bibfield  {journal} {\bibinfo
  {journal} {Nat. Mater.}\ }\textbf {\bibinfo {volume} {20}},\ \bibinfo {pages}
  {1601} (\bibinfo {year} {2021})}\BibitemShut {NoStop}%
\bibitem [{\citenamefont {Du}\ \emph {et~al.}(2021)\citenamefont {Du},
  \citenamefont {Lu},\ and\ \citenamefont {Xie}}]{Lu2021}%
  \BibitemOpen
  \bibfield  {author} {\bibinfo {author} {\bibfnamefont {Z.~Z.}\ \bibnamefont
  {Du}}, \bibinfo {author} {\bibfnamefont {H.-Z.}\ \bibnamefont {Lu}}, \ and\
  \bibinfo {author} {\bibfnamefont {X.}~\bibnamefont {Xie}},\ }\href {\doibase
  10.1038/s42254-021-00359-6} {\bibfield  {journal} {\bibinfo  {journal} {Nat.
  Rev. Phys.}\ }\textbf {\bibinfo {volume} {3}},\ \bibinfo {pages} {744}
  (\bibinfo {year} {2021})}\BibitemShut {NoStop}%
\bibitem [{\citenamefont {Sodemann}\ and\ \citenamefont {Fu}(2015)}]{Fu2015}%
  \BibitemOpen
  \bibfield  {author} {\bibinfo {author} {\bibfnamefont {I.}~\bibnamefont
  {Sodemann}}\ and\ \bibinfo {author} {\bibfnamefont {L.}~\bibnamefont {Fu}},\
  }\href {\doibase 10.1103/PhysRevLett.115.216806} {\bibfield  {journal}
  {\bibinfo  {journal} {Phys. Rev. Lett.}\ }\textbf {\bibinfo {volume} {115}},\
  \bibinfo {pages} {216806} (\bibinfo {year} {2015})}\BibitemShut {NoStop}%
\bibitem [{\citenamefont {Ma}\ \emph {et~al.}(2019)\citenamefont {Ma},
  \citenamefont {Xu}, \citenamefont {Shen}, \citenamefont {MacNeill},
  \citenamefont {Fatemi}, \citenamefont {Chang}, \citenamefont {Mier~Valdivia},
  \citenamefont {Wu}, \citenamefont {Du}, \citenamefont {Hsu}, \citenamefont
  {Fang}, \citenamefont {Gibson}, \citenamefont {Watanabe}, \citenamefont
  {Taniguchi}, \citenamefont {Cava}, \citenamefont {Kaxiras}, \citenamefont
  {Lu}, \citenamefont {Lin}, \citenamefont {Fu}, \citenamefont {Gedik},\ and\
  \citenamefont {Jarillo-Herrero}}]{Ma2019}%
  \BibitemOpen
  \bibfield  {author} {\bibinfo {author} {\bibfnamefont {Q.}~\bibnamefont
  {Ma}}, \bibinfo {author} {\bibfnamefont {S.-Y.}\ \bibnamefont {Xu}}, \bibinfo
  {author} {\bibfnamefont {H.}~\bibnamefont {Shen}}, \bibinfo {author}
  {\bibfnamefont {D.}~\bibnamefont {MacNeill}}, \bibinfo {author}
  {\bibfnamefont {V.}~\bibnamefont {Fatemi}}, \bibinfo {author} {\bibfnamefont
  {T.-R.}\ \bibnamefont {Chang}}, \bibinfo {author} {\bibfnamefont {A.~M.}\
  \bibnamefont {Mier~Valdivia}}, \bibinfo {author} {\bibfnamefont
  {S.}~\bibnamefont {Wu}}, \bibinfo {author} {\bibfnamefont {Z.}~\bibnamefont
  {Du}}, \bibinfo {author} {\bibfnamefont {C.-H.}\ \bibnamefont {Hsu}},
  \bibinfo {author} {\bibfnamefont {S.}~\bibnamefont {Fang}}, \bibinfo {author}
  {\bibfnamefont {Q.~D.}\ \bibnamefont {Gibson}}, \bibinfo {author}
  {\bibfnamefont {K.}~\bibnamefont {Watanabe}}, \bibinfo {author}
  {\bibfnamefont {T.}~\bibnamefont {Taniguchi}}, \bibinfo {author}
  {\bibfnamefont {R.~J.}\ \bibnamefont {Cava}}, \bibinfo {author}
  {\bibfnamefont {E.}~\bibnamefont {Kaxiras}}, \bibinfo {author} {\bibfnamefont
  {H.-Z.}\ \bibnamefont {Lu}}, \bibinfo {author} {\bibfnamefont
  {H.}~\bibnamefont {Lin}}, \bibinfo {author} {\bibfnamefont {L.}~\bibnamefont
  {Fu}}, \bibinfo {author} {\bibfnamefont {N.}~\bibnamefont {Gedik}}, \ and\
  \bibinfo {author} {\bibfnamefont {P.}~\bibnamefont {Jarillo-Herrero}},\
  }\href {\doibase 10.1038/s41586-018-0807-6} {\bibfield  {journal} {\bibinfo
  {journal} {Nature}\ }\textbf {\bibinfo {volume} {565}},\ \bibinfo {pages}
  {337} (\bibinfo {year} {2019})}\BibitemShut {NoStop}%
\bibitem [{\citenamefont {Kang}\ \emph {et~al.}(2019)\citenamefont {Kang},
  \citenamefont {Li}, \citenamefont {Sohn}, \citenamefont {Shan},\ and\
  \citenamefont {Mak}}]{Kang2019}%
  \BibitemOpen
  \bibfield  {author} {\bibinfo {author} {\bibfnamefont {K.}~\bibnamefont
  {Kang}}, \bibinfo {author} {\bibfnamefont {T.}~\bibnamefont {Li}}, \bibinfo
  {author} {\bibfnamefont {E.}~\bibnamefont {Sohn}}, \bibinfo {author}
  {\bibfnamefont {J.}~\bibnamefont {Shan}}, \ and\ \bibinfo {author}
  {\bibfnamefont {K.~F.}\ \bibnamefont {Mak}},\ }\href {\doibase
  10.1038/s41563-019-0294-7} {\bibfield  {journal} {\bibinfo  {journal} {Nat.
  Mater.}\ }\textbf {\bibinfo {volume} {18}},\ \bibinfo {pages} {324} (\bibinfo
  {year} {2019})}\BibitemShut {NoStop}%
\bibitem [{\citenamefont {Gao}\ \emph {et~al.}(2014)\citenamefont {Gao},
  \citenamefont {Yang},\ and\ \citenamefont {Niu}}]{Gao2014}%
  \BibitemOpen
  \bibfield  {author} {\bibinfo {author} {\bibfnamefont {Y.}~\bibnamefont
  {Gao}}, \bibinfo {author} {\bibfnamefont {S.~A.}\ \bibnamefont {Yang}}, \
  and\ \bibinfo {author} {\bibfnamefont {Q.}~\bibnamefont {Niu}},\ }\href
  {\doibase 10.1103/PhysRevLett.112.166601} {\bibfield  {journal} {\bibinfo
  {journal} {Phys. Rev. Lett.}\ }\textbf {\bibinfo {volume} {112}},\ \bibinfo
  {pages} {166601} (\bibinfo {year} {2014})}\BibitemShut {NoStop}%
\bibitem [{\citenamefont {Lai}\ \emph {et~al.}(2021)\citenamefont {Lai},
  \citenamefont {Liu}, \citenamefont {Zhang}, \citenamefont {Zhao},
  \citenamefont {Feng}, \citenamefont {Wang}, \citenamefont {Tang},
  \citenamefont {Liu}, \citenamefont {Novoselov}, \citenamefont {Yang},\ and\
  \citenamefont {Gao}}]{Lai2021}%
  \BibitemOpen
  \bibfield  {author} {\bibinfo {author} {\bibfnamefont {S.}~\bibnamefont
  {Lai}}, \bibinfo {author} {\bibfnamefont {H.}~\bibnamefont {Liu}}, \bibinfo
  {author} {\bibfnamefont {Z.}~\bibnamefont {Zhang}}, \bibinfo {author}
  {\bibfnamefont {J.}~\bibnamefont {Zhao}}, \bibinfo {author} {\bibfnamefont
  {X.}~\bibnamefont {Feng}}, \bibinfo {author} {\bibfnamefont {N.}~\bibnamefont
  {Wang}}, \bibinfo {author} {\bibfnamefont {C.}~\bibnamefont {Tang}}, \bibinfo
  {author} {\bibfnamefont {Y.}~\bibnamefont {Liu}}, \bibinfo {author}
  {\bibfnamefont {K.~S.}\ \bibnamefont {Novoselov}}, \bibinfo {author}
  {\bibfnamefont {S.~A.}\ \bibnamefont {Yang}}, \ and\ \bibinfo {author}
  {\bibfnamefont {W.-b.}\ \bibnamefont {Gao}},\ }\href {\doibase
  10.1038/s41565-021-00917-0} {\bibfield  {journal} {\bibinfo  {journal} {Nat.
  Nanotechnol.}\ }\textbf {\bibinfo {volume} {16}},\ \bibinfo {pages} {869}
  (\bibinfo {year} {2021})}\BibitemShut {NoStop}%
\bibitem [{\citenamefont {Liu}\ \emph {et~al.}(2022)\citenamefont {Liu},
  \citenamefont {Zhao}, \citenamefont {Huang}, \citenamefont {Feng},
  \citenamefont {Xiao}, \citenamefont {Wu}, \citenamefont {Lai}, \citenamefont
  {Gao},\ and\ \citenamefont {Yang}}]{Liu2022}%
  \BibitemOpen
  \bibfield  {author} {\bibinfo {author} {\bibfnamefont {H.}~\bibnamefont
  {Liu}}, \bibinfo {author} {\bibfnamefont {J.}~\bibnamefont {Zhao}}, \bibinfo
  {author} {\bibfnamefont {Y.-X.}\ \bibnamefont {Huang}}, \bibinfo {author}
  {\bibfnamefont {X.}~\bibnamefont {Feng}}, \bibinfo {author} {\bibfnamefont
  {C.}~\bibnamefont {Xiao}}, \bibinfo {author} {\bibfnamefont {W.}~\bibnamefont
  {Wu}}, \bibinfo {author} {\bibfnamefont {S.}~\bibnamefont {Lai}}, \bibinfo
  {author} {\bibfnamefont {W.-b.}\ \bibnamefont {Gao}}, \ and\ \bibinfo
  {author} {\bibfnamefont {S.~A.}\ \bibnamefont {Yang}},\ }\href {\doibase
  10.1103/PhysRevB.105.045118} {\bibfield  {journal} {\bibinfo  {journal}
  {Phys. Rev. B}\ }\textbf {\bibinfo {volume} {105}},\ \bibinfo {pages}
  {045118} (\bibinfo {year} {2022})}\BibitemShut {NoStop}%
\bibitem [{\citenamefont {Shao}\ \emph {et~al.}(2020)\citenamefont {Shao},
  \citenamefont {Zhang}, \citenamefont {Gurung}, \citenamefont {Yang},\ and\
  \citenamefont {Tsymbal}}]{Shao2020}%
  \BibitemOpen
  \bibfield  {author} {\bibinfo {author} {\bibfnamefont {D.-F.}\ \bibnamefont
  {Shao}}, \bibinfo {author} {\bibfnamefont {S.-H.}\ \bibnamefont {Zhang}},
  \bibinfo {author} {\bibfnamefont {G.}~\bibnamefont {Gurung}}, \bibinfo
  {author} {\bibfnamefont {W.}~\bibnamefont {Yang}}, \ and\ \bibinfo {author}
  {\bibfnamefont {E.~Y.}\ \bibnamefont {Tsymbal}},\ }\href {\doibase
  10.1103/PhysRevLett.124.067203} {\bibfield  {journal} {\bibinfo  {journal}
  {Phys. Rev. Lett.}\ }\textbf {\bibinfo {volume} {124}},\ \bibinfo {pages}
  {067203} (\bibinfo {year} {2020})}\BibitemShut {NoStop}%
\bibitem [{\citenamefont {Wang}\ \emph {et~al.}(2021)\citenamefont {Wang},
  \citenamefont {Gao},\ and\ \citenamefont {Xiao}}]{Wang2021}%
  \BibitemOpen
  \bibfield  {author} {\bibinfo {author} {\bibfnamefont {C.}~\bibnamefont
  {Wang}}, \bibinfo {author} {\bibfnamefont {Y.}~\bibnamefont {Gao}}, \ and\
  \bibinfo {author} {\bibfnamefont {D.}~\bibnamefont {Xiao}},\ }\href {\doibase
  10.1103/PhysRevLett.127.277201} {\bibfield  {journal} {\bibinfo  {journal}
  {Phys. Rev. Lett.}\ }\textbf {\bibinfo {volume} {127}},\ \bibinfo {pages}
  {277201} (\bibinfo {year} {2021})}\BibitemShut {NoStop}%
\bibitem [{\citenamefont {Liu}\ \emph {et~al.}(2021)\citenamefont {Liu},
  \citenamefont {Zhao}, \citenamefont {Huang}, \citenamefont {Wu},
  \citenamefont {Sheng}, \citenamefont {Xiao},\ and\ \citenamefont
  {Yang}}]{Liu2021}%
  \BibitemOpen
  \bibfield  {author} {\bibinfo {author} {\bibfnamefont {H.}~\bibnamefont
  {Liu}}, \bibinfo {author} {\bibfnamefont {J.}~\bibnamefont {Zhao}}, \bibinfo
  {author} {\bibfnamefont {Y.-X.}\ \bibnamefont {Huang}}, \bibinfo {author}
  {\bibfnamefont {W.}~\bibnamefont {Wu}}, \bibinfo {author} {\bibfnamefont
  {X.-L.}\ \bibnamefont {Sheng}}, \bibinfo {author} {\bibfnamefont
  {C.}~\bibnamefont {Xiao}}, \ and\ \bibinfo {author} {\bibfnamefont {S.~A.}\
  \bibnamefont {Yang}},\ }\href {\doibase 10.1103/PhysRevLett.127.277202}
  {\bibfield  {journal} {\bibinfo  {journal} {Phys. Rev. Lett.}\ }\textbf
  {\bibinfo {volume} {127}},\ \bibinfo {pages} {277202} (\bibinfo {year}
  {2021})}\BibitemShut {NoStop}%
\bibitem [{\citenamefont {Facio}\ \emph {et~al.}(2018)\citenamefont {Facio},
  \citenamefont {Efremov}, \citenamefont {Koepernik}, \citenamefont {You},
  \citenamefont {Sodemann},\ and\ \citenamefont {van~den Brink}}]{Facio2018}%
  \BibitemOpen
  \bibfield  {author} {\bibinfo {author} {\bibfnamefont {J.~I.}\ \bibnamefont
  {Facio}}, \bibinfo {author} {\bibfnamefont {D.}~\bibnamefont {Efremov}},
  \bibinfo {author} {\bibfnamefont {K.}~\bibnamefont {Koepernik}}, \bibinfo
  {author} {\bibfnamefont {J.-S.}\ \bibnamefont {You}}, \bibinfo {author}
  {\bibfnamefont {I.}~\bibnamefont {Sodemann}}, \ and\ \bibinfo {author}
  {\bibfnamefont {J.}~\bibnamefont {van~den Brink}},\ }\href {\doibase
  10.1103/PhysRevLett.121.246403} {\bibfield  {journal} {\bibinfo  {journal}
  {Phys. Rev. Lett.}\ }\textbf {\bibinfo {volume} {121}},\ \bibinfo {pages}
  {246403} (\bibinfo {year} {2018})}\BibitemShut {NoStop}%
\bibitem [{\citenamefont {Sinha}\ \emph {et~al.}(2022)\citenamefont {Sinha},
  \citenamefont {Adak}, \citenamefont {Chakraborty}, \citenamefont {Das},
  \citenamefont {Debnath}, \citenamefont {Sangani}, \citenamefont {Watanabe},
  \citenamefont {Taniguchi}, \citenamefont {Waghmare}, \citenamefont
  {Agarwal},\ and\ \citenamefont {Deshmukh}}]{Amit2022}%
  \BibitemOpen
  \bibfield  {author} {\bibinfo {author} {\bibfnamefont {S.}~\bibnamefont
  {Sinha}}, \bibinfo {author} {\bibfnamefont {P.~C.}\ \bibnamefont {Adak}},
  \bibinfo {author} {\bibfnamefont {A.}~\bibnamefont {Chakraborty}}, \bibinfo
  {author} {\bibfnamefont {K.}~\bibnamefont {Das}}, \bibinfo {author}
  {\bibfnamefont {K.}~\bibnamefont {Debnath}}, \bibinfo {author} {\bibfnamefont
  {L.~D.~V.}\ \bibnamefont {Sangani}}, \bibinfo {author} {\bibfnamefont
  {K.}~\bibnamefont {Watanabe}}, \bibinfo {author} {\bibfnamefont
  {T.}~\bibnamefont {Taniguchi}}, \bibinfo {author} {\bibfnamefont {U.~V.}\
  \bibnamefont {Waghmare}}, \bibinfo {author} {\bibfnamefont {A.}~\bibnamefont
  {Agarwal}}, \ and\ \bibinfo {author} {\bibfnamefont {M.~M.}\ \bibnamefont
  {Deshmukh}},\ }\href {\doibase 10.1038/s41567-022-01606-y} {\bibfield
  {journal} {\bibinfo  {journal} {Nat. Phys.}\ ,\ \bibinfo {pages} {1}}
  (\bibinfo {year} {2022})}\BibitemShut {NoStop}%
\bibitem [{\citenamefont {Awschalom}\ and\ \citenamefont
  {Samarth}(2009)}]{Review2009}%
  \BibitemOpen
  \bibfield  {author} {\bibinfo {author} {\bibfnamefont {D.}~\bibnamefont
  {Awschalom}}\ and\ \bibinfo {author} {\bibfnamefont {N.}~\bibnamefont
  {Samarth}},\ }\href {\doibase 10.1103/Physics.2.50} {\bibfield  {journal}
  {\bibinfo  {journal} {Physics}\ }\textbf {\bibinfo {volume} {2}},\ \bibinfo
  {pages} {50} (\bibinfo {year} {2009})}\BibitemShut {NoStop}%
\bibitem [{\citenamefont {Manchon}\ \emph {et~al.}(2019)\citenamefont
  {Manchon}, \citenamefont {\ifmmode~\check{Z}\else \v{Z}\fi{}elezn\'y},
  \citenamefont {Miron}, \citenamefont {Jungwirth}, \citenamefont {Sinova},
  \citenamefont {Thiaville}, \citenamefont {Garello},\ and\ \citenamefont
  {Gambardella}}]{Manchon2019}%
  \BibitemOpen
  \bibfield  {author} {\bibinfo {author} {\bibfnamefont {A.}~\bibnamefont
  {Manchon}}, \bibinfo {author} {\bibfnamefont {J.}~\bibnamefont
  {\ifmmode~\check{Z}\else \v{Z}\fi{}elezn\'y}}, \bibinfo {author}
  {\bibfnamefont {I.~M.}\ \bibnamefont {Miron}}, \bibinfo {author}
  {\bibfnamefont {T.}~\bibnamefont {Jungwirth}}, \bibinfo {author}
  {\bibfnamefont {J.}~\bibnamefont {Sinova}}, \bibinfo {author} {\bibfnamefont
  {A.}~\bibnamefont {Thiaville}}, \bibinfo {author} {\bibfnamefont
  {K.}~\bibnamefont {Garello}}, \ and\ \bibinfo {author} {\bibfnamefont
  {P.}~\bibnamefont {Gambardella}},\ }\href {\doibase
  10.1103/RevModPhys.91.035004} {\bibfield  {journal} {\bibinfo  {journal}
  {Rev. Mod. Phys.}\ }\textbf {\bibinfo {volume} {91}},\ \bibinfo {pages}
  {035004} (\bibinfo {year} {2019})}\BibitemShut {NoStop}%
\bibitem [{\citenamefont {Freimuth}\ \emph {et~al.}(2014)\citenamefont
  {Freimuth}, \citenamefont {Bl\"ugel},\ and\ \citenamefont
  {Mokrousov}}]{Freimuth2014}%
  \BibitemOpen
  \bibfield  {author} {\bibinfo {author} {\bibfnamefont {F.}~\bibnamefont
  {Freimuth}}, \bibinfo {author} {\bibfnamefont {S.}~\bibnamefont {Bl\"ugel}},
  \ and\ \bibinfo {author} {\bibfnamefont {Y.}~\bibnamefont {Mokrousov}},\
  }\href {\doibase 10.1103/PhysRevB.90.174423} {\bibfield  {journal} {\bibinfo
  {journal} {Phys. Rev. B}\ }\textbf {\bibinfo {volume} {90}},\ \bibinfo
  {pages} {174423} (\bibinfo {year} {2014})}\BibitemShut {NoStop}%
\bibitem [{\citenamefont {\ifmmode~\check{Z}\else \v{Z}\fi{}elezn\'y}\ \emph
  {et~al.}(2017)\citenamefont {\ifmmode~\check{Z}\else \v{Z}\fi{}elezn\'y},
  \citenamefont {Gao}, \citenamefont {Manchon}, \citenamefont {Freimuth},
  \citenamefont {Mokrousov}, \citenamefont {Zemen}, \citenamefont
  {Ma\ifmmode~\check{s}\else \v{s}\fi{}ek}, \citenamefont {Sinova},\ and\
  \citenamefont {Jungwirth}}]{Zelezny2017}%
  \BibitemOpen
  \bibfield  {author} {\bibinfo {author} {\bibfnamefont {J.}~\bibnamefont
  {\ifmmode~\check{Z}\else \v{Z}\fi{}elezn\'y}}, \bibinfo {author}
  {\bibfnamefont {H.}~\bibnamefont {Gao}}, \bibinfo {author} {\bibfnamefont
  {A.}~\bibnamefont {Manchon}}, \bibinfo {author} {\bibfnamefont
  {F.}~\bibnamefont {Freimuth}}, \bibinfo {author} {\bibfnamefont
  {Y.}~\bibnamefont {Mokrousov}}, \bibinfo {author} {\bibfnamefont
  {J.}~\bibnamefont {Zemen}}, \bibinfo {author} {\bibfnamefont
  {J.}~\bibnamefont {Ma\ifmmode~\check{s}\else \v{s}\fi{}ek}}, \bibinfo
  {author} {\bibfnamefont {J.}~\bibnamefont {Sinova}}, \ and\ \bibinfo {author}
  {\bibfnamefont {T.}~\bibnamefont {Jungwirth}},\ }\href {\doibase
  10.1103/PhysRevB.95.014403} {\bibfield  {journal} {\bibinfo  {journal} {Phys.
  Rev. B}\ }\textbf {\bibinfo {volume} {95}},\ \bibinfo {pages} {014403}
  (\bibinfo {year} {2017})}\BibitemShut {NoStop}%
\bibitem [{\citenamefont {Ivchenko}\ and\ \citenamefont
  {Pikus}(1978)}]{Pikus1978}%
  \BibitemOpen
  \bibfield  {author} {\bibinfo {author} {\bibfnamefont {E.~L.}\ \bibnamefont
  {Ivchenko}}\ and\ \bibinfo {author} {\bibfnamefont {G.~E.}\ \bibnamefont
  {Pikus}},\ }\href@noop {} {\bibfield  {journal} {\bibinfo  {journal} {JETP
  Lett.}\ }\textbf {\bibinfo {volume} {27}},\ \bibinfo {pages} {604} (\bibinfo
  {year} {1978})}\BibitemShut {NoStop}%
\bibitem [{\citenamefont {Aronov}\ and\ \citenamefont
  {Lyanda-Geller}(1989)}]{Aronov1989}%
  \BibitemOpen
  \bibfield  {author} {\bibinfo {author} {\bibfnamefont {A.~G.}\ \bibnamefont
  {Aronov}}\ and\ \bibinfo {author} {\bibfnamefont {Y.}~\bibnamefont
  {Lyanda-Geller}},\ }\href@noop {} {\bibfield  {journal} {\bibinfo  {journal}
  {JETP Lett.}\ }\textbf {\bibinfo {volume} {50}},\ \bibinfo {pages} {431}
  (\bibinfo {year} {1989})}\BibitemShut {NoStop}%
\bibitem [{\citenamefont {Edelstein}(1990)}]{Edelstein}%
  \BibitemOpen
  \bibfield  {author} {\bibinfo {author} {\bibfnamefont {Y.~M.}\ \bibnamefont
  {Edelstein}},\ }\href {\doibase 10.1016/0038-1098(90)90963-C} {\bibfield
  {journal} {\bibinfo  {journal} {Solid State Commun.}\ }\textbf {\bibinfo
  {volume} {73}},\ \bibinfo {pages} {233} (\bibinfo {year} {1990})}\BibitemShut
  {NoStop}%
\bibitem [{\citenamefont {Culcer}\ and\ \citenamefont
  {Winkler}(2007)}]{Culcer2007}%
  \BibitemOpen
  \bibfield  {author} {\bibinfo {author} {\bibfnamefont {D.}~\bibnamefont
  {Culcer}}\ and\ \bibinfo {author} {\bibfnamefont {R.}~\bibnamefont
  {Winkler}},\ }\href {\doibase 10.1103/PhysRevLett.99.226601} {\bibfield
  {journal} {\bibinfo  {journal} {Phys. Rev. Lett.}\ }\textbf {\bibinfo
  {volume} {99}},\ \bibinfo {pages} {226601} (\bibinfo {year}
  {2007})}\BibitemShut {NoStop}%
\bibitem [{\citenamefont {Chernyshov}\ \emph {et~al.}(2009)\citenamefont
  {Chernyshov}, \citenamefont {Overby}, \citenamefont {Liu}, \citenamefont
  {Furdyna}, \citenamefont {Lyanda-Geller},\ and\ \citenamefont
  {Rokhinson}}]{Geller2009}%
  \BibitemOpen
  \bibfield  {author} {\bibinfo {author} {\bibfnamefont {A.}~\bibnamefont
  {Chernyshov}}, \bibinfo {author} {\bibfnamefont {M.}~\bibnamefont {Overby}},
  \bibinfo {author} {\bibfnamefont {X.}~\bibnamefont {Liu}}, \bibinfo {author}
  {\bibfnamefont {J.~K.}\ \bibnamefont {Furdyna}}, \bibinfo {author}
  {\bibfnamefont {Y.}~\bibnamefont {Lyanda-Geller}}, \ and\ \bibinfo {author}
  {\bibfnamefont {L.~P.}\ \bibnamefont {Rokhinson}},\ }\href {\doibase
  10.1038/nphys1362} {\bibfield  {journal} {\bibinfo  {journal} {Nat. Phys.}\
  }\textbf {\bibinfo {volume} {5}},\ \bibinfo {pages} {656} (\bibinfo {year}
  {2009})}\BibitemShut {NoStop}%
\bibitem [{\citenamefont {Garate}\ and\ \citenamefont
  {MacDonald}(2009)}]{Garate2009}%
  \BibitemOpen
  \bibfield  {author} {\bibinfo {author} {\bibfnamefont {I.}~\bibnamefont
  {Garate}}\ and\ \bibinfo {author} {\bibfnamefont {A.~H.}\ \bibnamefont
  {MacDonald}},\ }\href {\doibase 10.1103/PhysRevB.80.134403} {\bibfield
  {journal} {\bibinfo  {journal} {Phys. Rev. B}\ }\textbf {\bibinfo {volume}
  {80}},\ \bibinfo {pages} {134403} (\bibinfo {year} {2009})}\BibitemShut
  {NoStop}%
\bibitem [{\citenamefont {Garate}\ and\ \citenamefont
  {Franz}(2010)}]{Franz2010}%
  \BibitemOpen
  \bibfield  {author} {\bibinfo {author} {\bibfnamefont {I.}~\bibnamefont
  {Garate}}\ and\ \bibinfo {author} {\bibfnamefont {M.}~\bibnamefont {Franz}},\
  }\href {\doibase 10.1103/PhysRevLett.104.146802} {\bibfield  {journal}
  {\bibinfo  {journal} {Phys. Rev. Lett.}\ }\textbf {\bibinfo {volume} {104}},\
  \bibinfo {pages} {146802} (\bibinfo {year} {2010})}\BibitemShut {NoStop}%
\bibitem [{\citenamefont {Miron}\ \emph {et~al.}(2010)\citenamefont {Miron},
  \citenamefont {Gaudin}, \citenamefont {Auffret}, \citenamefont {Rodmacq},
  \citenamefont {Schuhl}, \citenamefont {Pizzini}, \citenamefont {Vogel},\ and\
  \citenamefont {Gambardella}}]{Miron2010}%
  \BibitemOpen
  \bibfield  {author} {\bibinfo {author} {\bibfnamefont {I.~M.}\ \bibnamefont
  {Miron}}, \bibinfo {author} {\bibfnamefont {G.}~\bibnamefont {Gaudin}},
  \bibinfo {author} {\bibfnamefont {S.}~\bibnamefont {Auffret}}, \bibinfo
  {author} {\bibfnamefont {B.}~\bibnamefont {Rodmacq}}, \bibinfo {author}
  {\bibfnamefont {A.}~\bibnamefont {Schuhl}}, \bibinfo {author} {\bibfnamefont
  {S.}~\bibnamefont {Pizzini}}, \bibinfo {author} {\bibfnamefont
  {J.}~\bibnamefont {Vogel}}, \ and\ \bibinfo {author} {\bibfnamefont
  {P.}~\bibnamefont {Gambardella}},\ }\href {\doibase 10.1038/nmat2613}
  {\bibfield  {journal} {\bibinfo  {journal} {Nat. Mater.}\ }\textbf {\bibinfo
  {volume} {9}},\ \bibinfo {pages} {230} (\bibinfo {year} {2010})}\BibitemShut
  {NoStop}%
\bibitem [{\citenamefont {Fang}\ \emph {et~al.}(2011)\citenamefont {Fang},
  \citenamefont {Kurebayashi}, \citenamefont {Wunderlich}, \citenamefont
  {V\'yborn\'y}, \citenamefont {Z\^arbo}, \citenamefont {Campion},
  \citenamefont {Casiraghi}, \citenamefont {Gallagher}, \citenamefont
  {Jungwirth},\ and\ \citenamefont {Ferguson}}]{Vyborny2011}%
  \BibitemOpen
  \bibfield  {author} {\bibinfo {author} {\bibfnamefont {D.}~\bibnamefont
  {Fang}}, \bibinfo {author} {\bibfnamefont {H.}~\bibnamefont {Kurebayashi}},
  \bibinfo {author} {\bibfnamefont {J.}~\bibnamefont {Wunderlich}}, \bibinfo
  {author} {\bibfnamefont {K.}~\bibnamefont {V\'yborn\'y}}, \bibinfo {author}
  {\bibfnamefont {L.~P.}\ \bibnamefont {Z\^arbo}}, \bibinfo {author}
  {\bibfnamefont {R.~P.}\ \bibnamefont {Campion}}, \bibinfo {author}
  {\bibfnamefont {A.}~\bibnamefont {Casiraghi}}, \bibinfo {author}
  {\bibfnamefont {B.~L.}\ \bibnamefont {Gallagher}}, \bibinfo {author}
  {\bibfnamefont {T.}~\bibnamefont {Jungwirth}}, \ and\ \bibinfo {author}
  {\bibfnamefont {A.~J.}\ \bibnamefont {Ferguson}},\ }\href {\doibase
  10.1038/nnano.2011.68} {\bibfield  {journal} {\bibinfo  {journal} {Nat.
  Nanotechnol.}\ }\textbf {\bibinfo {volume} {6}},\ \bibinfo {pages} {413}
  (\bibinfo {year} {2011})}\BibitemShut {NoStop}%
\bibitem [{\citenamefont {Miron}\ \emph {et~al.}(2011)\citenamefont {Miron},
  \citenamefont {Garello}, \citenamefont {Gaudin}, \citenamefont {Zermatten},
  \citenamefont {Costache}, \citenamefont {Auffret}, \citenamefont {Bandiera},
  \citenamefont {Rodmacq}, \citenamefont {Schuhl},\ and\ \citenamefont
  {Gambardella}}]{Miron2011}%
  \BibitemOpen
  \bibfield  {author} {\bibinfo {author} {\bibfnamefont {I.~M.}\ \bibnamefont
  {Miron}}, \bibinfo {author} {\bibfnamefont {K.}~\bibnamefont {Garello}},
  \bibinfo {author} {\bibfnamefont {G.}~\bibnamefont {Gaudin}}, \bibinfo
  {author} {\bibfnamefont {P.-J.}\ \bibnamefont {Zermatten}}, \bibinfo {author}
  {\bibfnamefont {M.~V.}\ \bibnamefont {Costache}}, \bibinfo {author}
  {\bibfnamefont {S.}~\bibnamefont {Auffret}}, \bibinfo {author} {\bibfnamefont
  {S.}~\bibnamefont {Bandiera}}, \bibinfo {author} {\bibfnamefont
  {B.}~\bibnamefont {Rodmacq}}, \bibinfo {author} {\bibfnamefont
  {A.}~\bibnamefont {Schuhl}}, \ and\ \bibinfo {author} {\bibfnamefont
  {P.}~\bibnamefont {Gambardella}},\ }\href {\doibase 10.1038/nature10309}
  {\bibfield  {journal} {\bibinfo  {journal} {Nature}\ }\textbf {\bibinfo
  {volume} {476}},\ \bibinfo {pages} {189} (\bibinfo {year}
  {2011})}\BibitemShut {NoStop}%
\bibitem [{\citenamefont {Liu}\ \emph {et~al.}(2012)\citenamefont {Liu},
  \citenamefont {Pai}, \citenamefont {Li}, \citenamefont {Tseng}, \citenamefont
  {Ralph},\ and\ \citenamefont {Buhrman}}]{Liu2012}%
  \BibitemOpen
  \bibfield  {author} {\bibinfo {author} {\bibfnamefont {L.}~\bibnamefont
  {Liu}}, \bibinfo {author} {\bibfnamefont {C.-F.}\ \bibnamefont {Pai}},
  \bibinfo {author} {\bibfnamefont {Y.}~\bibnamefont {Li}}, \bibinfo {author}
  {\bibfnamefont {H.~W.}\ \bibnamefont {Tseng}}, \bibinfo {author}
  {\bibfnamefont {D.~C.}\ \bibnamefont {Ralph}}, \ and\ \bibinfo {author}
  {\bibfnamefont {R.~A.}\ \bibnamefont {Buhrman}},\ }\href {\doibase
  10.1126/science.1218197} {\bibfield  {journal} {\bibinfo  {journal}
  {Science}\ }\textbf {\bibinfo {volume} {336}},\ \bibinfo {pages} {555}
  (\bibinfo {year} {2012})}\BibitemShut {NoStop}%
\bibitem [{\citenamefont {Garello}\ \emph {et~al.}(2013)\citenamefont
  {Garello}, \citenamefont {Miron}, \citenamefont {Avci}, \citenamefont
  {Freimuth}, \citenamefont {Mokrousov}, \citenamefont {Bl{\"u}gel},
  \citenamefont {Auffret}, \citenamefont {Boulle}, \citenamefont {Gaudin},\
  and\ \citenamefont {Gambardella}}]{Garello2013}%
  \BibitemOpen
  \bibfield  {author} {\bibinfo {author} {\bibfnamefont {K.}~\bibnamefont
  {Garello}}, \bibinfo {author} {\bibfnamefont {I.~M.}\ \bibnamefont {Miron}},
  \bibinfo {author} {\bibfnamefont {C.~O.}\ \bibnamefont {Avci}}, \bibinfo
  {author} {\bibfnamefont {F.}~\bibnamefont {Freimuth}}, \bibinfo {author}
  {\bibfnamefont {Y.}~\bibnamefont {Mokrousov}}, \bibinfo {author}
  {\bibfnamefont {S.}~\bibnamefont {Bl{\"u}gel}}, \bibinfo {author}
  {\bibfnamefont {S.}~\bibnamefont {Auffret}}, \bibinfo {author} {\bibfnamefont
  {O.}~\bibnamefont {Boulle}}, \bibinfo {author} {\bibfnamefont
  {G.}~\bibnamefont {Gaudin}}, \ and\ \bibinfo {author} {\bibfnamefont
  {P.}~\bibnamefont {Gambardella}},\ }\href {\doibase 10.1038/nnano.2013.145}
  {\bibfield  {journal} {\bibinfo  {journal} {Nat. Nanotechnol.}\ }\textbf
  {\bibinfo {volume} {8}},\ \bibinfo {pages} {587} (\bibinfo {year}
  {2013})}\BibitemShut {NoStop}%
\bibitem [{\citenamefont {Kurebayashi}\ \emph {et~al.}(2014)\citenamefont
  {Kurebayashi}, \citenamefont {Sinova}, \citenamefont {Fang}, \citenamefont
  {Irvine}, \citenamefont {Skinner}, \citenamefont {Wunderlich}, \citenamefont
  {Nov{\'a}k}, \citenamefont {Campion}, \citenamefont {Gallagher},
  \citenamefont {Vehstedt}, \citenamefont {Z\^arbo}, \citenamefont
  {V\'yborn\'y}, \citenamefont {Ferguson},\ and\ \citenamefont
  {Jungwirth}}]{Kurebayashi2014}%
  \BibitemOpen
  \bibfield  {author} {\bibinfo {author} {\bibfnamefont {H.}~\bibnamefont
  {Kurebayashi}}, \bibinfo {author} {\bibfnamefont {J.}~\bibnamefont {Sinova}},
  \bibinfo {author} {\bibfnamefont {D.}~\bibnamefont {Fang}}, \bibinfo {author}
  {\bibfnamefont {A.}~\bibnamefont {Irvine}}, \bibinfo {author} {\bibfnamefont
  {T.~D.}\ \bibnamefont {Skinner}}, \bibinfo {author} {\bibfnamefont
  {J.}~\bibnamefont {Wunderlich}}, \bibinfo {author} {\bibfnamefont
  {V.}~\bibnamefont {Nov{\'a}k}}, \bibinfo {author} {\bibfnamefont {R.~P.}\
  \bibnamefont {Campion}}, \bibinfo {author} {\bibfnamefont {B.~L.}\
  \bibnamefont {Gallagher}}, \bibinfo {author} {\bibfnamefont {E.~K.}\
  \bibnamefont {Vehstedt}}, \bibinfo {author} {\bibfnamefont {L.~P.}\
  \bibnamefont {Z\^arbo}}, \bibinfo {author} {\bibfnamefont {K.}~\bibnamefont
  {V\'yborn\'y}}, \bibinfo {author} {\bibfnamefont {A.~J.}\ \bibnamefont
  {Ferguson}}, \ and\ \bibinfo {author} {\bibfnamefont {T.}~\bibnamefont
  {Jungwirth}},\ }\href {\doibase 10.1038/nnano.2014.15} {\bibfield  {journal}
  {\bibinfo  {journal} {Nat. Nanotechnol.}\ }\textbf {\bibinfo {volume} {9}},\
  \bibinfo {pages} {211} (\bibinfo {year} {2014})}\BibitemShut {NoStop}%
\bibitem [{\citenamefont {Ciccarelli}\ \emph {et~al.}(2016)\citenamefont
  {Ciccarelli}, \citenamefont {Anderson}, \citenamefont {Tshitoyan},
  \citenamefont {Ferguson}, \citenamefont {Gerhard}, \citenamefont {Gould},
  \citenamefont {Molenkamp}, \citenamefont {Gayles}, \citenamefont
  {{\v{Z}}elezn{\`y}}, \citenamefont {{\v{S}}mejkal}, \citenamefont {Yuan},
  \citenamefont {Sinova}, \citenamefont {Freimuth},\ and\ \citenamefont
  {Jungwirth}}]{Jungwirth2016}%
  \BibitemOpen
  \bibfield  {author} {\bibinfo {author} {\bibfnamefont {C.}~\bibnamefont
  {Ciccarelli}}, \bibinfo {author} {\bibfnamefont {L.}~\bibnamefont
  {Anderson}}, \bibinfo {author} {\bibfnamefont {V.}~\bibnamefont {Tshitoyan}},
  \bibinfo {author} {\bibfnamefont {A.~J.}\ \bibnamefont {Ferguson}}, \bibinfo
  {author} {\bibfnamefont {F.}~\bibnamefont {Gerhard}}, \bibinfo {author}
  {\bibfnamefont {C.}~\bibnamefont {Gould}}, \bibinfo {author} {\bibfnamefont
  {L.~W.}\ \bibnamefont {Molenkamp}}, \bibinfo {author} {\bibfnamefont
  {J.}~\bibnamefont {Gayles}}, \bibinfo {author} {\bibfnamefont
  {J.}~\bibnamefont {{\v{Z}}elezn{\`y}}}, \bibinfo {author} {\bibfnamefont
  {L.}~\bibnamefont {{\v{S}}mejkal}}, \bibinfo {author} {\bibfnamefont
  {Z.}~\bibnamefont {Yuan}}, \bibinfo {author} {\bibfnamefont {J.}~\bibnamefont
  {Sinova}}, \bibinfo {author} {\bibfnamefont {F.}~\bibnamefont {Freimuth}}, \
  and\ \bibinfo {author} {\bibfnamefont {T.}~\bibnamefont {Jungwirth}},\ }\href
  {\doibase 10.1038/nphys3772} {\bibfield  {journal} {\bibinfo  {journal} {Nat.
  Phys.}\ }\textbf {\bibinfo {volume} {12}},\ \bibinfo {pages} {855} (\bibinfo
  {year} {2016})}\BibitemShut {NoStop}%
\bibitem [{\citenamefont {Xiao}\ \emph {et~al.}(2022)\citenamefont {Xiao},
  \citenamefont {Liu}, \citenamefont {Wu}, \citenamefont {Wang}, \citenamefont
  {Niu},\ and\ \citenamefont {Yang}}]{Xiao2022NLISOT}%
  \BibitemOpen
  \bibfield  {author} {\bibinfo {author} {\bibfnamefont {C.}~\bibnamefont
  {Xiao}}, \bibinfo {author} {\bibfnamefont {H.}~\bibnamefont {Liu}}, \bibinfo
  {author} {\bibfnamefont {W.}~\bibnamefont {Wu}}, \bibinfo {author}
  {\bibfnamefont {H.}~\bibnamefont {Wang}}, \bibinfo {author} {\bibfnamefont
  {Q.}~\bibnamefont {Niu}}, \ and\ \bibinfo {author} {\bibfnamefont {S.~A.}\
  \bibnamefont {Yang}},\ }\href {\doibase 10.1103/PhysRevLett.129.086602}
  {\bibfield  {journal} {\bibinfo  {journal} {Phys. Rev. Lett.}\ }\textbf
  {\bibinfo {volume} {129}},\ \bibinfo {pages} {086602} (\bibinfo {year}
  {2022})}\BibitemShut {NoStop}%
\bibitem [{sup()}]{supp}%
  \BibitemOpen
  \href@noop {} {}\bibinfo {howpublished} {See Supplemental Material for the
  detailed matrix forms of $\alpha_{a(bc)}^{\text{even}}$ in all nonmagnetic
  and ferromagnetic point groups that forbid the linear CISP, the comparison of
  symmetry enforced forms of $\alpha^{\text{even}}$ and $\alpha^{\text{odd}}$
  in such ferromagnetic point groups, and the computational details in concrete
  materials.}\BibitemShut {Stop}%
\bibitem [{\citenamefont {Dong}\ \emph {et~al.}(2020)\citenamefont {Dong},
  \citenamefont {Xiao}, \citenamefont {Xiong},\ and\ \citenamefont
  {Niu}}]{Dong2020}%
  \BibitemOpen
  \bibfield  {author} {\bibinfo {author} {\bibfnamefont {L.}~\bibnamefont
  {Dong}}, \bibinfo {author} {\bibfnamefont {C.}~\bibnamefont {Xiao}}, \bibinfo
  {author} {\bibfnamefont {B.}~\bibnamefont {Xiong}}, \ and\ \bibinfo {author}
  {\bibfnamefont {Q.}~\bibnamefont {Niu}},\ }\href {\doibase
  10.1103/PhysRevLett.124.066601} {\bibfield  {journal} {\bibinfo  {journal}
  {Phys. Rev. Lett.}\ }\textbf {\bibinfo {volume} {124}},\ \bibinfo {pages}
  {066601} (\bibinfo {year} {2020})}\BibitemShut {NoStop}%
\bibitem [{\citenamefont {Xiao}\ \emph {et~al.}(2010)\citenamefont {Xiao},
  \citenamefont {Chang},\ and\ \citenamefont {Niu}}]{Xiao2010}%
  \BibitemOpen
  \bibfield  {author} {\bibinfo {author} {\bibfnamefont {D.}~\bibnamefont
  {Xiao}}, \bibinfo {author} {\bibfnamefont {M.-C.}\ \bibnamefont {Chang}}, \
  and\ \bibinfo {author} {\bibfnamefont {Q.}~\bibnamefont {Niu}},\ }\href
  {\doibase 10.1103/RevModPhys.82.1959} {\bibfield  {journal} {\bibinfo
  {journal} {Rev. Mod. Phys.}\ }\textbf {\bibinfo {volume} {82}},\ \bibinfo
  {pages} {1959} (\bibinfo {year} {2010})}\BibitemShut {NoStop}%
\bibitem [{\citenamefont {Chang}\ and\ \citenamefont {Niu}(1995)}]{Chang1995}%
  \BibitemOpen
  \bibfield  {author} {\bibinfo {author} {\bibfnamefont {M.-C.}\ \bibnamefont
  {Chang}}\ and\ \bibinfo {author} {\bibfnamefont {Q.}~\bibnamefont {Niu}},\
  }\href {\doibase 10.1103/PhysRevLett.75.1348} {\bibfield  {journal} {\bibinfo
   {journal} {Phys. Rev. Lett.}\ }\textbf {\bibinfo {volume} {75}},\ \bibinfo
  {pages} {1348} (\bibinfo {year} {1995})}\BibitemShut {NoStop}%
\bibitem [{\citenamefont {Sundaram}\ and\ \citenamefont
  {Niu}(1999)}]{Sundaram1999}%
  \BibitemOpen
  \bibfield  {author} {\bibinfo {author} {\bibfnamefont {G.}~\bibnamefont
  {Sundaram}}\ and\ \bibinfo {author} {\bibfnamefont {Q.}~\bibnamefont {Niu}},\
  }\href {\doibase 10.1103/PhysRevB.59.14915} {\bibfield  {journal} {\bibinfo
  {journal} {Phys. Rev. B}\ }\textbf {\bibinfo {volume} {59}},\ \bibinfo
  {pages} {14915} (\bibinfo {year} {1999})}\BibitemShut {NoStop}%
\bibitem [{\citenamefont {Kane}\ and\ \citenamefont {Mele}(2005)}]{Kane2005}%
  \BibitemOpen
  \bibfield  {author} {\bibinfo {author} {\bibfnamefont {C.~L.}\ \bibnamefont
  {Kane}}\ and\ \bibinfo {author} {\bibfnamefont {E.~J.}\ \bibnamefont
  {Mele}},\ }\href {\doibase 10.1103/PhysRevLett.95.226801} {\bibfield
  {journal} {\bibinfo  {journal} {Phys. Rev. Lett.}\ }\textbf {\bibinfo
  {volume} {95}},\ \bibinfo {pages} {226801} (\bibinfo {year}
  {2005})}\BibitemShut {NoStop}%
\bibitem [{\citenamefont {Liu}\ \emph {et~al.}(2011{\natexlab{a}})\citenamefont
  {Liu}, \citenamefont {Feng},\ and\ \citenamefont {Yao}}]{Yao2011PRL}%
  \BibitemOpen
  \bibfield  {author} {\bibinfo {author} {\bibfnamefont {C.-C.}\ \bibnamefont
  {Liu}}, \bibinfo {author} {\bibfnamefont {W.}~\bibnamefont {Feng}}, \ and\
  \bibinfo {author} {\bibfnamefont {Y.}~\bibnamefont {Yao}},\ }\href {\doibase
  10.1103/PhysRevLett.107.076802} {\bibfield  {journal} {\bibinfo  {journal}
  {Phys. Rev. Lett.}\ }\textbf {\bibinfo {volume} {107}},\ \bibinfo {pages}
  {076802} (\bibinfo {year} {2011}{\natexlab{a}})}\BibitemShut {NoStop}%
\bibitem [{\citenamefont {Liu}\ \emph {et~al.}(2011{\natexlab{b}})\citenamefont
  {Liu}, \citenamefont {Jiang},\ and\ \citenamefont {Yao}}]{Yao2011PRB}%
  \BibitemOpen
  \bibfield  {author} {\bibinfo {author} {\bibfnamefont {C.-C.}\ \bibnamefont
  {Liu}}, \bibinfo {author} {\bibfnamefont {H.}~\bibnamefont {Jiang}}, \ and\
  \bibinfo {author} {\bibfnamefont {Y.}~\bibnamefont {Yao}},\ }\href {\doibase
  10.1103/PhysRevB.84.195430} {\bibfield  {journal} {\bibinfo  {journal} {Phys.
  Rev. B}\ }\textbf {\bibinfo {volume} {84}},\ \bibinfo {pages} {195430}
  (\bibinfo {year} {2011}{\natexlab{b}})}\BibitemShut {NoStop}%
\bibitem [{Kit()}]{Kittel1996}%
  \BibitemOpen
  \href@noop {} {}\bibinfo {howpublished} {C. Kittel, {\it Introduction to
  Solid State Physics} (7th edition) (John Wiley and Sons, New York,
  1996).}\BibitemShut {Stop}%
\bibitem [{\citenamefont {Olejn{\'\i}k}\ \emph {et~al.}(2018)\citenamefont
  {Olejn{\'\i}k}, \citenamefont {Seifert}, \citenamefont {Ka{\v{s}}par},
  \citenamefont {Nov{\'a}k}, \citenamefont {Wadley}, \citenamefont {Campion},
  \citenamefont {Baumgartner}, \citenamefont {Gambardella}, \citenamefont
  {N{\v{e}}mec}, \citenamefont {Wunderlich}, \citenamefont {Sinova},
  \citenamefont {Ku{\v{z}}el}, \citenamefont {M{\"u}ller}, \citenamefont
  {Kampfrath},\ and\ \citenamefont {Jungwirth}}]{Jungwirth2018}%
  \BibitemOpen
  \bibfield  {author} {\bibinfo {author} {\bibfnamefont {K.}~\bibnamefont
  {Olejn{\'\i}k}}, \bibinfo {author} {\bibfnamefont {T.}~\bibnamefont
  {Seifert}}, \bibinfo {author} {\bibfnamefont {Z.}~\bibnamefont
  {Ka{\v{s}}par}}, \bibinfo {author} {\bibfnamefont {V.}~\bibnamefont
  {Nov{\'a}k}}, \bibinfo {author} {\bibfnamefont {P.}~\bibnamefont {Wadley}},
  \bibinfo {author} {\bibfnamefont {R.~P.}\ \bibnamefont {Campion}}, \bibinfo
  {author} {\bibfnamefont {M.}~\bibnamefont {Baumgartner}}, \bibinfo {author}
  {\bibfnamefont {P.}~\bibnamefont {Gambardella}}, \bibinfo {author}
  {\bibfnamefont {P.}~\bibnamefont {N{\v{e}}mec}}, \bibinfo {author}
  {\bibfnamefont {J.}~\bibnamefont {Wunderlich}}, \bibinfo {author}
  {\bibfnamefont {J.}~\bibnamefont {Sinova}}, \bibinfo {author} {\bibfnamefont
  {P.}~\bibnamefont {Ku{\v{z}}el}}, \bibinfo {author} {\bibfnamefont
  {M.}~\bibnamefont {M{\"u}ller}}, \bibinfo {author} {\bibfnamefont
  {T.}~\bibnamefont {Kampfrath}}, \ and\ \bibinfo {author} {\bibfnamefont
  {T.}~\bibnamefont {Jungwirth}},\ }\href {\doibase 10.1126/sciadv.aar3566}
  {\bibfield  {journal} {\bibinfo  {journal} {Sci. Adv.}\ }\textbf {\bibinfo
  {volume} {4}},\ \bibinfo {pages} {eaar3566} (\bibinfo {year}
  {2018})}\BibitemShut {NoStop}%
\bibitem [{\citenamefont {Zhou}\ \emph {et~al.}(2019)\citenamefont {Zhou},
  \citenamefont {Chen}, \citenamefont {Zhang}, \citenamefont {Li},
  \citenamefont {Shi}, \citenamefont {Sun}, \citenamefont {Saleem},
  \citenamefont {You}, \citenamefont {Pan},\ and\ \citenamefont
  {Song}}]{Song2019}%
  \BibitemOpen
  \bibfield  {author} {\bibinfo {author} {\bibfnamefont {X.~F.}\ \bibnamefont
  {Zhou}}, \bibinfo {author} {\bibfnamefont {X.~Z.}\ \bibnamefont {Chen}},
  \bibinfo {author} {\bibfnamefont {J.}~\bibnamefont {Zhang}}, \bibinfo
  {author} {\bibfnamefont {F.}~\bibnamefont {Li}}, \bibinfo {author}
  {\bibfnamefont {G.~Y.}\ \bibnamefont {Shi}}, \bibinfo {author} {\bibfnamefont
  {Y.~M.}\ \bibnamefont {Sun}}, \bibinfo {author} {\bibfnamefont {M.~S.}\
  \bibnamefont {Saleem}}, \bibinfo {author} {\bibfnamefont {Y.~F.}\
  \bibnamefont {You}}, \bibinfo {author} {\bibfnamefont {F.}~\bibnamefont
  {Pan}}, \ and\ \bibinfo {author} {\bibfnamefont {C.}~\bibnamefont {Song}},\
  }\href {\doibase 10.1103/PhysRevApplied.11.054030} {\bibfield  {journal}
  {\bibinfo  {journal} {Phys. Rev. Appl.}\ }\textbf {\bibinfo {volume} {11}},\
  \bibinfo {pages} {054030} (\bibinfo {year} {2019})}\BibitemShut {NoStop}%
\bibitem [{\citenamefont {Freimuth}\ \emph {et~al.}(2010)\citenamefont
  {Freimuth}, \citenamefont {Bl\"ugel},\ and\ \citenamefont
  {Mokrousov}}]{Freimuth2010}%
  \BibitemOpen
  \bibfield  {author} {\bibinfo {author} {\bibfnamefont {F.}~\bibnamefont
  {Freimuth}}, \bibinfo {author} {\bibfnamefont {S.}~\bibnamefont {Bl\"ugel}},
  \ and\ \bibinfo {author} {\bibfnamefont {Y.}~\bibnamefont {Mokrousov}},\
  }\href {\doibase 10.1103/PhysRevLett.105.246602} {\bibfield  {journal}
  {\bibinfo  {journal} {Phys. Rev. Lett.}\ }\textbf {\bibinfo {volume} {105}},\
  \bibinfo {pages} {246602} (\bibinfo {year} {2010})}\BibitemShut {NoStop}%
\bibitem [{\citenamefont {Kato}\ \emph {et~al.}(2004)\citenamefont {Kato},
  \citenamefont {Myers}, \citenamefont {Gossard},\ and\ \citenamefont
  {Awschalom}}]{Kato2004}%
  \BibitemOpen
  \bibfield  {author} {\bibinfo {author} {\bibfnamefont {Y.~K.}\ \bibnamefont
  {Kato}}, \bibinfo {author} {\bibfnamefont {R.~C.}\ \bibnamefont {Myers}},
  \bibinfo {author} {\bibfnamefont {A.~C.}\ \bibnamefont {Gossard}}, \ and\
  \bibinfo {author} {\bibfnamefont {D.~D.}\ \bibnamefont {Awschalom}},\ }\href
  {\doibase 10.1103/PhysRevLett.93.176601} {\bibfield  {journal} {\bibinfo
  {journal} {Phys. Rev. Lett.}\ }\textbf {\bibinfo {volume} {93}},\ \bibinfo
  {pages} {176601} (\bibinfo {year} {2004})}\BibitemShut {NoStop}%
\bibitem [{\citenamefont {Stern}\ \emph {et~al.}(2006)\citenamefont {Stern},
  \citenamefont {Ghosh}, \citenamefont {Xiang}, \citenamefont {Zhu},
  \citenamefont {Samarth},\ and\ \citenamefont {Awschalom}}]{Stern2006}%
  \BibitemOpen
  \bibfield  {author} {\bibinfo {author} {\bibfnamefont {N.~P.}\ \bibnamefont
  {Stern}}, \bibinfo {author} {\bibfnamefont {S.}~\bibnamefont {Ghosh}},
  \bibinfo {author} {\bibfnamefont {G.}~\bibnamefont {Xiang}}, \bibinfo
  {author} {\bibfnamefont {M.}~\bibnamefont {Zhu}}, \bibinfo {author}
  {\bibfnamefont {N.}~\bibnamefont {Samarth}}, \ and\ \bibinfo {author}
  {\bibfnamefont {D.~D.}\ \bibnamefont {Awschalom}},\ }\href {\doibase
  10.1103/PhysRevLett.97.126603} {\bibfield  {journal} {\bibinfo  {journal}
  {Phys. Rev. Lett.}\ }\textbf {\bibinfo {volume} {97}},\ \bibinfo {pages}
  {126603} (\bibinfo {year} {2006})}\BibitemShut {NoStop}%
\bibitem [{\citenamefont {Chen}\ \emph {et~al.}(2017)\citenamefont {Chen},
  \citenamefont {Pai}, \citenamefont {Chan}, \citenamefont {Takayama},
  \citenamefont {Xu}, \citenamefont {Karn}, \citenamefont {Hasegawa},
  \citenamefont {Chou}, \citenamefont {Mo}, \citenamefont {Fedorov},\ and\
  \citenamefont {Chiang}}]{Chiang2017}%
  \BibitemOpen
  \bibfield  {author} {\bibinfo {author} {\bibfnamefont {P.}~\bibnamefont
  {Chen}}, \bibinfo {author} {\bibfnamefont {W.~W.}\ \bibnamefont {Pai}},
  \bibinfo {author} {\bibfnamefont {Y.-H.}\ \bibnamefont {Chan}}, \bibinfo
  {author} {\bibfnamefont {A.}~\bibnamefont {Takayama}}, \bibinfo {author}
  {\bibfnamefont {C.-Z.}\ \bibnamefont {Xu}}, \bibinfo {author} {\bibfnamefont
  {A.}~\bibnamefont {Karn}}, \bibinfo {author} {\bibfnamefont {S.}~\bibnamefont
  {Hasegawa}}, \bibinfo {author} {\bibfnamefont {M.-Y.}\ \bibnamefont {Chou}},
  \bibinfo {author} {\bibfnamefont {S.-K.}\ \bibnamefont {Mo}}, \bibinfo
  {author} {\bibfnamefont {A.-V.}\ \bibnamefont {Fedorov}}, \ and\ \bibinfo
  {author} {\bibfnamefont {T.-C.}\ \bibnamefont {Chiang}},\ }\href {\doibase
  10.1038/s41467-017-00641-1} {\bibfield  {journal} {\bibinfo  {journal} {Nat.
  Commun.}\ }\textbf {\bibinfo {volume} {8}},\ \bibinfo {pages} {1} (\bibinfo
  {year} {2017})}\BibitemShut {NoStop}%
\bibitem [{\citenamefont {O'Hara}\ \emph {et~al.}(2018)\citenamefont {O'Hara},
  \citenamefont {Zhu}, \citenamefont {Trout}, \citenamefont {Ahmed},
  \citenamefont {Luo}, \citenamefont {Lee}, \citenamefont {Brenner},
  \citenamefont {Rajan}, \citenamefont {Gupta}, \citenamefont {McComb},\ and\
  \citenamefont {Kawakami}}]{Kawakami2018}%
  \BibitemOpen
  \bibfield  {author} {\bibinfo {author} {\bibfnamefont {D.~J.}\ \bibnamefont
  {O'Hara}}, \bibinfo {author} {\bibfnamefont {T.}~\bibnamefont {Zhu}},
  \bibinfo {author} {\bibfnamefont {A.~H.}\ \bibnamefont {Trout}}, \bibinfo
  {author} {\bibfnamefont {A.~S.}\ \bibnamefont {Ahmed}}, \bibinfo {author}
  {\bibfnamefont {Y.~K.}\ \bibnamefont {Luo}}, \bibinfo {author} {\bibfnamefont
  {C.~H.}\ \bibnamefont {Lee}}, \bibinfo {author} {\bibfnamefont {M.~R.}\
  \bibnamefont {Brenner}}, \bibinfo {author} {\bibfnamefont {S.}~\bibnamefont
  {Rajan}}, \bibinfo {author} {\bibfnamefont {J.~A.}\ \bibnamefont {Gupta}},
  \bibinfo {author} {\bibfnamefont {D.~W.}\ \bibnamefont {McComb}}, \ and\
  \bibinfo {author} {\bibfnamefont {R.~K.}\ \bibnamefont {Kawakami}},\ }\href
  {\doibase doi: 10.1021/acs.nanolett.8b00683} {\bibfield  {journal} {\bibinfo
  {journal} {Nano Lett.}\ }\textbf {\bibinfo {volume} {18}},\ \bibinfo {pages}
  {3125} (\bibinfo {year} {2018})}\BibitemShut {NoStop}%
\bibitem [{\citenamefont {Vanherck}\ \emph {et~al.}(2020)\citenamefont
  {Vanherck}, \citenamefont {Bacaksiz}, \citenamefont {Sor{\'e}e},
  \citenamefont {Milo{\v{s}}evi{\'c}},\ and\ \citenamefont
  {Magnus}}]{Magnus2020}%
  \BibitemOpen
  \bibfield  {author} {\bibinfo {author} {\bibfnamefont {J.}~\bibnamefont
  {Vanherck}}, \bibinfo {author} {\bibfnamefont {C.}~\bibnamefont {Bacaksiz}},
  \bibinfo {author} {\bibfnamefont {B.}~\bibnamefont {Sor{\'e}e}}, \bibinfo
  {author} {\bibfnamefont {M.~V.}\ \bibnamefont {Milo{\v{s}}evi{\'c}}}, \ and\
  \bibinfo {author} {\bibfnamefont {W.}~\bibnamefont {Magnus}},\ }\href
  {\doibase 10.1063/5.0015619} {\bibfield  {journal} {\bibinfo  {journal}
  {Appl. Phys. Lett.}\ }\textbf {\bibinfo {volume} {117}},\ \bibinfo {pages}
  {052401} (\bibinfo {year} {2020})}\BibitemShut {NoStop}%
\bibitem [{\citenamefont {Chen}\ \emph {et~al.}(2010)\citenamefont {Chen},
  \citenamefont {Qin}, \citenamefont {Yang}, \citenamefont {Liu}, \citenamefont
  {Guan}, \citenamefont {Qu}, \citenamefont {Zhang}, \citenamefont {Shi},
  \citenamefont {Xie}, \citenamefont {Yang}, \citenamefont {Wu}, \citenamefont
  {Li},\ and\ \citenamefont {Lu}}]{Xie2010}%
  \BibitemOpen
  \bibfield  {author} {\bibinfo {author} {\bibfnamefont {J.}~\bibnamefont
  {Chen}}, \bibinfo {author} {\bibfnamefont {H.~J.}\ \bibnamefont {Qin}},
  \bibinfo {author} {\bibfnamefont {F.}~\bibnamefont {Yang}}, \bibinfo {author}
  {\bibfnamefont {J.}~\bibnamefont {Liu}}, \bibinfo {author} {\bibfnamefont
  {T.}~\bibnamefont {Guan}}, \bibinfo {author} {\bibfnamefont {F.~M.}\
  \bibnamefont {Qu}}, \bibinfo {author} {\bibfnamefont {G.~H.}\ \bibnamefont
  {Zhang}}, \bibinfo {author} {\bibfnamefont {J.~R.}\ \bibnamefont {Shi}},
  \bibinfo {author} {\bibfnamefont {X.~C.}\ \bibnamefont {Xie}}, \bibinfo
  {author} {\bibfnamefont {C.~L.}\ \bibnamefont {Yang}}, \bibinfo {author}
  {\bibfnamefont {K.~H.}\ \bibnamefont {Wu}}, \bibinfo {author} {\bibfnamefont
  {Y.~Q.}\ \bibnamefont {Li}}, \ and\ \bibinfo {author} {\bibfnamefont
  {L.}~\bibnamefont {Lu}},\ }\href {\doibase 10.1103/PhysRevLett.105.176602}
  {\bibfield  {journal} {\bibinfo  {journal} {Phys. Rev. Lett.}\ }\textbf
  {\bibinfo {volume} {105}},\ \bibinfo {pages} {176602} (\bibinfo {year}
  {2010})}\BibitemShut {NoStop}%
\bibitem [{\citenamefont {Du}\ \emph {et~al.}(2019)\citenamefont {Du},
  \citenamefont {Wang}, \citenamefont {Li}, \citenamefont {Lu},\ and\
  \citenamefont {Xie}}]{Du2019}%
  \BibitemOpen
  \bibfield  {author} {\bibinfo {author} {\bibfnamefont {Z.~Z.}\ \bibnamefont
  {Du}}, \bibinfo {author} {\bibfnamefont {C.~M.}\ \bibnamefont {Wang}},
  \bibinfo {author} {\bibfnamefont {S.}~\bibnamefont {Li}}, \bibinfo {author}
  {\bibfnamefont {H.-Z.}\ \bibnamefont {Lu}}, \ and\ \bibinfo {author}
  {\bibfnamefont {X.~C.}\ \bibnamefont {Xie}},\ }\href {\doibase
  10.1038/s41467-019-10941-3} {\bibfield  {journal} {\bibinfo  {journal} {Nat.
  Commun.}\ }\textbf {\bibinfo {volume} {10}},\ \bibinfo {pages} {3047}
  (\bibinfo {year} {2019})}\BibitemShut {NoStop}%
\bibitem [{\citenamefont {Xiao}\ \emph {et~al.}(2019)\citenamefont {Xiao},
  \citenamefont {Du},\ and\ \citenamefont {Niu}}]{Xiao2019NLHE}%
  \BibitemOpen
  \bibfield  {author} {\bibinfo {author} {\bibfnamefont {C.}~\bibnamefont
  {Xiao}}, \bibinfo {author} {\bibfnamefont {Z.~Z.}\ \bibnamefont {Du}}, \ and\
  \bibinfo {author} {\bibfnamefont {Q.}~\bibnamefont {Niu}},\ }\href {\doibase
  10.1103/PhysRevB.100.165422} {\bibfield  {journal} {\bibinfo  {journal}
  {Phys. Rev. B}\ }\textbf {\bibinfo {volume} {100}},\ \bibinfo {pages}
  {165422} (\bibinfo {year} {2019})}\BibitemShut {NoStop}%
\bibitem [{\citenamefont {\ifmmode~\check{Z}\else \v{Z}\fi{}elezn\'y}\ \emph
  {et~al.}(2014)\citenamefont {\ifmmode~\check{Z}\else \v{Z}\fi{}elezn\'y},
  \citenamefont {Gao}, \citenamefont {V\'yborn\'y}, \citenamefont {Zemen},
  \citenamefont {Ma\ifmmode~\check{s}\else \v{s}\fi{}ek}, \citenamefont
  {Manchon}, \citenamefont {Wunderlich}, \citenamefont {Sinova},\ and\
  \citenamefont {Jungwirth}}]{Zelezny2014}%
  \BibitemOpen
  \bibfield  {author} {\bibinfo {author} {\bibfnamefont {J.}~\bibnamefont
  {\ifmmode~\check{Z}\else \v{Z}\fi{}elezn\'y}}, \bibinfo {author}
  {\bibfnamefont {H.}~\bibnamefont {Gao}}, \bibinfo {author} {\bibfnamefont
  {K.}~\bibnamefont {V\'yborn\'y}}, \bibinfo {author} {\bibfnamefont
  {J.}~\bibnamefont {Zemen}}, \bibinfo {author} {\bibfnamefont
  {J.}~\bibnamefont {Ma\ifmmode~\check{s}\else \v{s}\fi{}ek}}, \bibinfo
  {author} {\bibfnamefont {A.}~\bibnamefont {Manchon}}, \bibinfo {author}
  {\bibfnamefont {J.}~\bibnamefont {Wunderlich}}, \bibinfo {author}
  {\bibfnamefont {J.}~\bibnamefont {Sinova}}, \ and\ \bibinfo {author}
  {\bibfnamefont {T.}~\bibnamefont {Jungwirth}},\ }\href {\doibase
  10.1103/PhysRevLett.113.157201} {\bibfield  {journal} {\bibinfo  {journal}
  {Phys. Rev. Lett.}\ }\textbf {\bibinfo {volume} {113}},\ \bibinfo {pages}
  {157201} (\bibinfo {year} {2014})}\BibitemShut {NoStop}%
\end{thebibliography}%

\end{document}